\documentclass{article}

\usepackage{algorithm}
\usepackage[margin=30mm,includehead,includefoot]{geometry}
\usepackage[noend]{algpseudocode}
\usepackage{amsmath,amssymb,amsthm,amscd,color,comment}
\usepackage{autobreak}
\usepackage{color}
\usepackage{xcolor}
\usepackage{url}
\usepackage{blkarray}
\usepackage{jheppub}
\usepackage{empheq}
\usepackage{graphicx}
\usepackage{booktabs, multirow}
\usepackage{subfig}
\usepackage{amsmath}
\usepackage{adjustbox}

\usepackage[nottoc]{tocbibind}

\theoremstyle{plain}

\theoremstyle{definition}

\usepackage{xpatch}
\makeatletter
\AtBeginDocument{\xpatchcmd{\@thm}{\thm@headpunct{.}}{\thm@headpunct{}}{}{}}
\makeatother

\newcommand{\id}{\mathrm{d}}

\makeatletter
\newlength{\apb@width}
\newcommand{\autoparbox}[2][c]{\settowidth{\apb@width}{#2}\parbox[#1]{\apb@width}{#2}}

\makeatother

\title{The Loop-by-Loop Baikov Representation - \\ Strategies and Implementation}

\author[a]{Hjalte Frellesvig}

\affiliation[a]{Niels Bohr International Academy, University of Copenhagen,
Blegdamsvej 17, 2100 K{\o}benhavn {\O}, Denmark}

\emailAdd{hjalte.frellesvig@nbi.ku.dk}

\abstract{
In this paper, we discuss the Baikov representation of Feynman integrals in its standard and loop-by-loop variants. The Baikov representation is a parametric representation, which has as its defining feature the fact that the integration variables are the propagators of the Feynman integral. For the loop-by-loop Baikov representation, we discuss in detail a strategy for how to make an optimal parametrization which is one that minimizes the number of extra integration variables that have to be introduced for a given Feynman integral. Furthermore, we present a \texttt{Mathematica} implementation, named \texttt{BaikovPackage}, that is able to generate the Baikov representation in its standard and loop-by-loop varieties. We also discuss some subtleties and open problems regarding Baikov representations.
}

\begin{document}
\maketitle

\nocite{Frellesvig:2017aai}

\section{Introduction}
\label{sec:intro}

Feynman integrals are an important tool in modern physics. They are essential for computing higher order perturbative contributions to scattering cross sections and other observables in quantum field theory, and as such necessary for computing theoretical predictions for, for instance, the ongoing particle scattering experiments at the Large Hadron Collider. Feynman integrals have also found uses in other fields of physics, such as the investigation of gravitational waves emitted by colliding black holes and other heavy astronomical objects.

As Feynman integrals are prone to divergences they need to be regulated. In this paper we will solely consider the usual choice of dimensional regularization, in which the Feynman integral will be given as an integral over a number of $d$-dimensional loop momenta, where $d$ will be considered a non-integer; specifically we will take $d = d_{\text{spacetime}} - 2 \epsilon$. Evaluating, or even conceptualizing, an integral over a non-integer number of variables is not straight forward. For that reason it is often useful to express it in a \textit{parametric representation}, by which we will mean a representation where the Feynman integral is given as an integral over an integer number of scalar integration variables.

The by far most well-known parametric representation of Feynman integrals is the Feynman parameter representation. It is introduced in most text-books on quantum field theory, and historically many Feynman integrals have been computed and investigated with this representation as a starting point. Along with the Schwinger representation and the Lee--Pomeransky representation~\cite{Lee:2013hzt}, the Feynman parameter representation has the Feynman integral expressed with an integrand that contains two so-called Symanzik polynomials named $\mathcal{U}$ and $\mathcal{F}$, and together these representation could be said to form a family of Feynman integral representations. See e.g. refs.~\cite{Weinzierl:2022eaz, Smirnov:2012gma} for book length introductions.

In this paper we will discuss a different family of representations, namely variations over the \textit{Baikov representation}. This parametrization was originally introduced in refs.~\cite{Baikov:1996cd, Baikov:1996iu, Baikov:2005nv} by P. A. Baikov after whom it is named. It was then further used, discussed, investigated, or developed in e.g. refs.~\cite{Lee:2010wea, Grozin:2011mt, Smirnov:2012gma, Larsen:2015ped, Zhang:2016kfo, Harley:2017qut, Bosma:2017ens, Frellesvig:2017aai, Mastrolia:2018uzb, Frellesvig:2019kgj, Frellesvig:2021vdl, Weinzierl:2022eaz, Chen:2022lzr, Jiang:2023qnl, Jiang:2023oyq}. We will discuss two variants called, respectively, the \textit{standard} Baikov representation and the \textit{loop-by-loop}~\cite{Frellesvig:2017aai} Baikov representation. We will introduce these representations in detail in the following section, but briefly stated in the loop-by-loop representation the integral gets Baikov parametrized one loop at a time, whereas in the standard Baikov representation the whole integral gets parametrized simultaneously.

The Baikov representation is usually derived from the momentum representation through a series of variable changes, as discussed e.g. in app.~\ref{app:derivation}. Stopping at intermediate steps in that derivation, for instance corresponding to eq.~\eqref{eq:intstep1} where the Feynman integral is expressed as an integral over parallel and radial components of the loop momenta can, however, also yield useful results~\cite{Cutkosky:singu, Caron-Huot:2021xqj, Vergu:2023rqz}. This is sometimes known as the Cutkosky representation~\cite{Cutkosky:singu, Vergu:2023rqz}, and it may be considered a close cousin to the family of Baikov representations.\\

The defining feature of Baikov representations is the fact that the integration variables are the propagators of the Feynman integral. This has the advantage\footnote{In fairness we should also mention something that the Baikov representation is not recommended for, which is actually evaluating the Feynman integrals. Compared to the Feynman parameter representation and its relatives, the Baikov representation usually has more integration variables, more complicated defining polynomials, a much more complicated integration region, and it lacks the projective property of the Feynman parameter representation that allows for the use of the Cheng--Wu theorem.} that performing \textit{generalized cuts}~\cite{Britto:2004nc, Cachazo:2008vp, Britto:2024mna} becomes a trivial operation. Generalized cuts (also known as ``generalized unitarity cuts'') are traditionally discussed as replacing the cut propagator with a delta function, but are more properly understood as a deformation of the integration contour into one encircling the poles formed by the cut propagators going on shell. Since in the Baikov representation the integration variables are the propagators, a generalized cut just becomes a trivial residue operation in the Baikov variables that correspond to the cut propagators.

It is this ease with which generalized cuts are performed in the Baikov representation, that is the key to many of its most prominent use cases. Particularly useful are \textit{maximal cuts}, which we here will define as generalized cuts that fix as many degrees of freedom as possible. Prominently for polylogarithmic Feynman integrals, \textit{pure integrals}~\cite{Arkani-Hamed:2010pyv} by definition have their maximal cut being constant in the kinematics. Pure Feynman integrals have the property~\cite{Dlapa:2021qsl} that their \textit{differential equations}~\cite{KOTIKOV1991158} have a particularly nice, \textit{canonical}, $\epsilon$-factorized form~\cite{Henn:2013pwa}, a fact that it used for most state-of-the-art Feynman integral computations.
For non-polylogarithmic Feynman integrals, maximal cuts can be used to reveal the underlying geometric structure~\cite{Laporta:2004rb, Primo:2016ebd, Primo:2017ipr, Frellesvig:2023bbf}, such as an elliptic curve or a given Calabi--Yau manifold, over which the Feynman integral can be expressed as an iterated integral. Also in these cases this knowledge may be used to bring the differential equations obeyed by the Feynman integral into an $\epsilon$-factorized form~\cite{Adams:2018bsn, Broedel:2019kmn, Frellesvig:2021hkr, Duhr:2022dxb}. There are cases in which the Baikov-based approach to these geometric structures seems to disagree with the result obtained using different approaches~\cite{Frellesvig:2021vdl, Marzucca:2023gto}, and in those cases it tends to be the Baikov-based approach that leads to the correct result.

Even without heavy utilization of generalized cuts, the various types of linear relations between Feynman integrals may be fruitfully investigated from the perspective of the Baikov representation. This is the case for Integration-By-Part (IBP) identities which become more transparent from that point of view (the Baikov representation is being used by the recently published IBP-program NeatIBP~\cite{Wu:2023upw}). It also holds for the differential equations relating Feynman integrals, as well as for dimension shift relations, the investigation of which was one of the first uses of the Baikov representation~\cite{Lee:2010wea}.

Another prominent feature of the Baikov representation is that it allows for the direct identification of Feynman integrals with \textit{Aomoto Gel'fand integrals} also known as \textit{generalized Euler integrals} among other names~\cite{Matsubara-Heo:2023ylc}. In particular this identification allows for a direct mapping of concepts from the world of Feynman integrals into the mathematical framework of \textit{twisted cohomology}~\cite{Mastrolia:2018uzb, Frellesvig:2019uqt}. This can be used to formalize the vector space structure~\cite{Smirnov:2010hn, Lee:2013hzt, Frellesvig:2019uqt} of a Feynman integral family, and, prominently, to enable the use of \textit{intersection theory}\footnote{In this context intersection theory refers to the intersection theory of twisted cocycles. That branch of mathematics has been developed since the 1990s~\cite{cho1995, matsumoto1998}, it was introduced in physics in 2017~\cite{Mizera:2017cqs, Mizera:2017rqa}, and shortly thereafter the connection to Feynman integrals was realized~\cite{Mastrolia:2018uzb}.}~\cite{Mastrolia:2018uzb, Frellesvig:2019uqt} to derive relations between Feynman integrals, an approach which is emerging as a promising alternative~\cite{Brunello:2023rpq, Brunello:2024tqf} to traditional IBP based methods.

Furthermore the Baikov representation has recently been shown to have a natural connection~\cite{Caron-Huot:2024brh} to the \textit{Landau singularity} structure of Feynman integrals, something that had previously been investigated mainly using Feynman parameters. Relatedly the \textit{symbol letter} structure of polylogarithmic Feynman integrals, may also be derived using insights generated from the Baikov representation~\cite{Chen:2023kgw, Jiang:2023qnl, Jiang:2024eaj} and its relation to intersection theory.\\

The purpose of this paper is threefold. The first goal is to summarize and discuss known facts about the Baikov representation in its standard and loop-by-loop varieties. That is done in this introduction, along with sections \ref{sec:baikov} and \ref{sec:perspectives} and the four appendices. The second goal is to discuss strategies for how to approach the loop-by-loop Baikov representation for a given Feynman integral. This is done in sec.~\ref{sec:strategies} under the name of ``the graphical approach'', which has not previously been presented in the literature. Lastly the third goal is to present and formally publish an implementation\footnote{There are other public implementations of the Baikov representation. One is \texttt{Baikov.m}, an implementation of the standard Baikov representation published with ref.~\cite{Frellesvig:2017aai}. Another is the package \texttt{BaikovAll.wl} published with ref.~\cite{Jiang:2023qnl} which implements the recursive approach to loop-by-loop Baikov discussed in app.~\ref{app:recursive}. Yet another is the implementation of standard Baikov as the function \texttt{Prepare} inside \texttt{NeatIBP}~\cite{Wu:2023upw}.} of the Baikov representation as a \texttt{Mathematica} package named \texttt{BaikovPackage}. That package (which can be downloaded at \url{https://github.com/HjalteFrellesvig/BaikovPackage}) has been under development since 2018 and has been used for numerous research projects over that time, including refs.~\cite{Frellesvig:2019kgj, Frellesvig:2019uqt, Bonciani:2019jyb, Frellesvig:2019byn, Frellesvig:2020qot, Frellesvig:2021vdl, Frellesvig:2021hkr, Chestnov:2022xsy, Frellesvig:2023iwr, Frellesvig:2023bbf, Brunello:2023rpq, Frellesvig:2024swj, Frellesvig:2024zph, Frellesvig:2024rea} by the present author with various collaborators, along with (at least) refs.~\cite{Brunello:2023fef, Brunello:2024ibk, Brunello:2024tqf, Crisanti:2024onv, Duhr:2024bzt, Benincasa:2024ptf} by other authors.\\

The structure of the paper is as follows: In sec.~\ref{sec:baikov} we will introduce the Baikov representation in its two main variants. In sec.~\ref{sec:strategies} we will discuss the above-mentioned graphical approach for how to best perform a loop-by-loop Baikov parametrization of a given Feynman integral. In sec.~\ref{sec:BaikovPackage} we will describe the \texttt{Mathematica} implementation of the Baikov representation, named \texttt{BaikovPackage}, that is published alongside this paper. In sec.~\ref{sec:perspectives} we will discuss some open problems and loose ends connected to the Baikov representation. Lastly the paper contains four appendices. App.~\ref{app:derivation} contains the derivation of the Baikov representation in its standard and loop-by-loop variants. App.~\ref{app:cannot} goes over some types of integrals for which the Baikov representation cannot be applied directly and discusses what to do in such cases. App.~\ref{app:remaining} discusses how to make a loop-by-loop parametrization of Feynman integrals with numerator factors that seem to worsen the external momentum dependence of the individual loops, and lastly app.~\ref{app:recursive} discusses an alternative, recursive, approach to deriving the loop-by-loop Baikov representation.

\section{Baikov representations}
\label{sec:baikov}

In this section we will discuss two versions of the Baikov representation, \textit{standard} and \textit{loop-by-loop}. The derivation\footnote{It should be noted that the derivation of the Baikov representation performed in app.~\ref{app:derivation} is done in Euclidean space. Continuing the expression back to Minkowski space will introduce some factors of $i$ that are not present in the expressions in this paper. Furthermore the expressions here may have an ambiguity in the over-all sign due to the wedge-product between the integration variables. Thus the expressions for the Baikov representation given here, in particular eqs.~\eqref{eq:standard} and \eqref{eq:loop-by-loop}, may be considered ambiguous up to over-all factors of $i$. For more discussion of this, see refs.~\cite{Weinzierl:2022eaz, Harley:2017qut}.} of the expressions given here can be found in many places in the literature~\cite{Grozin:2011mt, Frellesvig:2017aai, Mastrolia:2018uzb, Frellesvig:2021vdl, Weinzierl:2022eaz}, and it is summarized in app.~\ref{app:derivation}.

The starting point will be a Feynman integral in momentum representation, where it is given as\footnote{In this section the propagators are denoted by $\rho_i$, but elsewhere in the paper we will use the notation $P_i$. Likewise the propagator powers $a_i$ will elsewhere be denoted $\nu_i$. We hope that this is not a source of confusion.}
\begin{align}
I &= \int \frac{\id^d k_1}{i \pi^{d/2}} \cdots \frac{\id^d k_L}{i \pi^{d/2}} \frac{N(k)}{\rho_1(k)^{a_1} \cdots \rho_P(k)^{a_P}}
\label{eq:momentumrep}
\end{align}
Here we have $d$ being the spacetime dimensionality in dimensional regularization, $L$ being the number of loops, $k_i$ the loop momenta, $P$ the number of propagators, $\rho_i(k)$ the propagators themselves, $a_i$ the powers of those propagators, and $N(k)$ a generic scalar numerator factor.

\subsection{The standard Baikov representation}
\label{sec:standard}

In the standard Baikov representation (also referred to as the ``democratic approach''~\cite{Weinzierl:2022eaz}), a Feynman integral is given by
\begin{align}
\boxed{
\; I \, = \, \frac{\mathcal{J} \; (-i)^L \, \pi^{(L-n)/2} \, \mathcal{E}^{(E-d+1)/2}}{\prod_{l=1}^L \Gamma \big( (d{+}1{-}E{-}l)/2 \big)} \int_{\mathcal{C}} \frac{ x_{P{+}1}^{-a_{P{+}1}} \cdots x_{n}^{-a_{n}} }{x_1^{a_1} \cdots x_{P}^{a_{P}}} \mathcal{B}^{(d{-}E{-}L{-}1)/2} \, \id^{n} x \;
}
\label{eq:standard}
\end{align}
Here $x_i$ are the scalar integration variables called \textbf{Baikov variables}, with the important property that the Baikov variables \textit{are} the propagators, i.e. $x_i = \rho_i$ for $i \leq P$.
Furthermore $E$ is\footnote{Please note that in this paper $E$ counts the number of independent \textit{external} momenta, and not the number of \textit{edges} as is the case elsewhere in the Feynman integral literature.} the number of independent external momenta, and $n$ is the total number of Baikov variables. In this case $n = n_{\text{std}}$ where
\begin{align}
n_{\text{std}} = L(L+1)/2 + EL
\label{eq:nstd}
\end{align}
counts the different scalar products that can be formed which involve the loop momenta, i.e. either of the form $k_i \cdot k_j$ or $k_i \cdot p_j$ where $p_i$ denotes the independent momenta external to the diagram.
The two polynomials $\mathcal{E}$ and $\mathcal{B}$ are defined as Gram determinants as 
\begin{align}
\mathcal{E} = \det \! \big( G(p_1,\ldots,p_E) \big) \;, \quad\qquad \mathcal{B} = \det \! \big( G(k_1,\ldots,k_L,p_1,\ldots,p_E) \big)
\label{eq:EandB}
\end{align}
where the Gram matrix $G(\{q\})$ is the matrix formed by all the scalar products $G_{ij} = q_i \cdot q_j$. Of these $\mathcal{B}$ is called the \textbf{Baikov polynomial}, while $\mathcal{E}$ could be called the \textbf{external Baikov polynomial}. The factor $\mathcal{J}$ is the Jacobian of a variable change from the independent scalar products to the Baikov variables. Its exact expression depends on exactly how the propagators are defined, but usually $\mathcal{J} = \pm 2^{L-n}$.
Lastly for the case of one loop, the integration region\footnote{With the integration region given by eq.~\eqref{eq:contourstandard}, the integral can only converge when the power $\gamma$ to which $\mathcal{B}$ is raised, obeys $\gamma > -1$. For other values the integral will have to be analytically continued. One way of doing that is to replace the integration region of eq.~\eqref{eq:contourstandard} with a multivariate generalization of the \textit{Pochhammer contour}~\cite{aomoto2011theory, yoshida2013hypergeometric, Matsubara-Heo:2023ylc}. This is a genuine cycle (or higher dimensional contour) embedded in $\mathbb{C}^n$ (thus justifying the use of the symbol $\mathcal{C}$), and it reduces to eq.~\eqref{eq:contourstandard} when $\gamma > -1$.} $\mathcal{C}$ is given by
\begin{align}
\mathcal{C} &= \left\{ x \in \mathbb{R}^n \,:\, \frac{\mathcal{B}(x)}{\mathcal{E}} > 0 \right\}
\label{eq:contourstandard}
\end{align}
i.e. the integration should be done over the parts of Baikov parameter space in which $\mathcal{B}/\mathcal{E}$ is positive. In the multi-loop case, the integration region is as for the loop-by-loop representation that we will discuss in eq.~\eqref{eq:contourlbl}.

For most Feynman integrals beyond one loop, there are fewer propagators than the $n$ given by eq.~\eqref{eq:nstd}. In that case the set of \textit{actual} or \textit{physical} propagators (written in the denominator of eq.~\eqref{eq:standard}) has to be supplemented by a set of \textit{extra} propagators (written in the numerator of eq.~\eqref{eq:standard}) that has no direct equivalent in eq.~\eqref{eq:momentumrep}. This has to be done in such a way that the total number of propagators (or rather propagator-type objects) equals $n$ and where furthermore the equation system relating the propagators to the scalar products involving the loop momenta has to be invertible. In other words we need 
\begin{align}
N_{\text{extra}} = n - P
\label{eq:Nextra}
\end{align}
such extra propagators.

If the original set of propagators is not linearly independent to begin with, applying the Baikov representation directly is not possible. What to do in such cases is discussed in detail in app.~\ref{app:cannot}.

We could have written eq.~\eqref{eq:standard} with a generic numerator factor $N(x)$, corresponding to re-expressing the $N(k)$ of eq.~\eqref{eq:momentumrep} in terms of the Baikov variables. Each monomial in that $N(x)$ may, however, be expressed on the form of the integrand of eq.~\eqref{eq:standard}, and thus an integral with a generic numerator can be written as a sum of terms of the form of eq.~\eqref{eq:standard}.

Most parametric representations of Feynman integrals, such as the Feynman parameter representation, will have a number of integration variables that equals the number of propagators. Yet the number of Baikov variables given by eq.~\eqref{eq:nstd} can be much larger since it grows quadratically with the number of loops. That motivates the search for modifications of the Baikov representation that allow for a smaller value of $N_{\text{extra}}$, and such a modification is given by the loop-by-loop Baikov representation.

\subsection{The loop-by-loop Baikov representation}
\label{sec:loop-by-loop}

The loop-by-loop representation may be obtained by applying the representation of eq.~\eqref{eq:standard} to each loop at a time, as opposed to the whole integral at once. The loop-by-loop Baikov representation is defined up to an ordering of the loops, corresponding to the order in which the loops get parametrized. For an $L$-loop integral we will index the loops with the variable $l$ that goes from $1$ to $L$, such that the loop with index $l$ is the $l$th loop to be parametrized. For a given loop $l'$ we will refer to loops with $l<l'$ as \textit{lower} loops, and to those with $l>l'$ as \textit{higher} loops. The loop-by-loop Baikov representation was originally introduced in ref.~\cite{Frellesvig:2017aai}, but most of the following expressions were first given explicitly in ref.~\cite{Frellesvig:2021vem}.

Inserting $L=1$ in eq.~\eqref{eq:standard} we obtain
\begin{align}
I_{\text{one-loop}} &= \frac{\mathcal{J}_{1} \, (-i) \, \pi^{-E/2} \, \mathcal{E}^{(E-d+1)/2}}{\Gamma \big( (d-E)/2 \big)} \int_{\mathcal{C}} \frac{ \mathcal{B}^{(d{-}E{-}2)/2} }{x_1^{a_1} \cdots x_{E{+}1}^{a_{E{+}1}}} \, \id^{E+1} x
\label{eq:oneloop}
\end{align}
and applying this expression recursively, as discussed in detail in app.~\ref{app:derivation}, gives the loop-by-loop Baikov representation
\begin{align}
\boxed{
\; I \, = \, \frac{\mathcal{J} \; (-i)^L \, \pi^{(L - n)/2}}{\prod_{l=1}^L \Gamma \big( (d-E_l)/2 \big)} \int_{\mathcal{C}} \frac{ x_{P{+}1}^{-a_{P{+}1}} \cdots x_n^{-a_n} }{x_1^{a_1} \cdots x_{P}^{a_{P}}} \left( \prod_{l=1}^L \mathcal{E}_l^{(E_l-d+1)/2} \, \mathcal{B}_l^{(d-E_l-2)/2} \right) \id^n x \; }
\label{eq:loop-by-loop}
\end{align}
As before $L$ is the number of loops, and $P$ the number of propagators. $E_l$ is the number of independent momenta external to loop number $l$ after having integrated out the lower loops (much more explanation of this is to follow in sec.~\ref{sec:strategies}.) The number of integration variables is again denoted $n$, but now it is given by $n = n_{\text{lbl}}$ where
\begin{align}
n_{\text{lbl}} = L + \sum_{l=1}^{L} E_l
\label{eq:nlbl}
\end{align}
Each loop now comes with two polynomials each, which are given by
\begin{align}
\mathcal{E}_l = \det \! \big( G(q_1,\ldots,q_{E_l}) \big) \;, \quad\qquad \mathcal{B}_l = \det \! \big( G(k_l,q_1,\ldots,q_{E_l}) \big)
\end{align}
where the $q_i$ are the $E_l$ momenta external to the $l$th loop. These momenta may be given by external momenta $p$, loop momenta of higher loops $k$, or combinations thereof. Note that $\mathcal{E}_L$ always will be constant in the Baikov variables, so only $2L-1$ of the polynomials have to be under the integral sign. The Jacobian $\mathcal{J}$ is again that of the variable change from the scalar products to the Baikov variables, and it is usually given by $\mathcal{J} = \pm 2^{L-n}$.
Lastly the integration region $\mathcal{C}$ is defined by
\begin{align}
\mathcal{C} &= \left\{ x \in \mathbb{R}^n \,:\, \bigwedge_{l=1}^L \frac{\mathcal{B}_l(x)}{\mathcal{E}_l(x)} > 0 \right\}
\label{eq:contourlbl}
\end{align}
i.e. the integration should be done over the parts of Baikov parameter space that correspond to the intersection of the integration regions of the $L$ individual loops.

Whenever the numerator of the original integral of eq.~\eqref{eq:momentumrep} contains a dot-product that is not present in the propagators of a loop-by-loop parametrization of that integral, it might seem as if that loop-by-loop parametrization would not be applicable. That would, however, be a wrong conclusion; the loop-by-loop representation can be adapted to also include such cases. How to do that is the subject of app.~\ref{app:remaining}.

If all loops have dependence on all external momenta and all loop momenta of higher loops we have $E_l = E + L - l$. Inserting this in eq.~\eqref{eq:nlbl} gives $n_{\text{lbl}} = n_{\text{std}}$, and in fact the two representations are equivalent in that case. This property can be induced artificially by introducing extra propagators such that all scalar products between all momenta are present, and the introduction of the fact that the powers of these extra propagators are zero can be postponed to the end. This is how the standard Baikov representation is derived in app.~\ref{app:derivation}. Yet for all other cases than the one where all loops depend on all external momenta and all loop momenta of higher loops, we have $n_{\text{lbl}} < n_{\text{std}}$. This implies that the number of extra variables that has to be introduced in order to achieve a valid parametrization (i.e. the $N_{\text{extra}}$ of eq.~\eqref{eq:Nextra}) is smaller by the same amount in the loop-by-loop case. It is also possible to derive the loop-by-loop Baikov representation starting from the standard representation by recursively integrating the extra variables out, and this is discussed in detail in app.~\ref{app:recursive}.

We will now discuss how to pick the loop order and the extra propagators, in such a way that $N_{\text{extra}}$ becomes as small as possible.

\section{Strategies for loop-by-loop Baikov representation}
\label{sec:strategies}

In this section we will discuss strategies for how to pick a loop ordering for a loop-by-loop parametrization, in such a way as to minimize the number of extra propagators needed. We will do this by discussing a specific example in detail, and then deduce a general strategy that we will name ``the graphical approach''. That strategy will be independent of the internal masses of the diagrams, and almost\footnote{The counter example is Feynman integrals with massless two-point kinematics which have to be treated differently. This is discussed in app.~\ref{app:masslesstwopoint}.} independent of the external masses. For that reason we will, without loss of generality, mostly discuss cases with massless internal and external kinematics.

\subsection{The box-triangle}
\label{sec:boxtriangle}

\begin{figure}[h]
\centering
\vspace{-2mm}
\includegraphics[width=5cm]{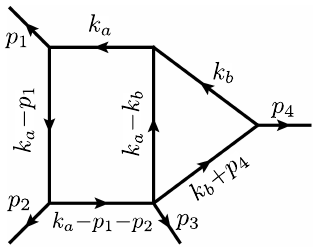}
\vspace{-1mm}
\caption{The box-triangle diagram that will be our main example.}
\label{fig:boxtri}
\end{figure}

Let us start by considering the box-triangle Feynman integral shown in fig.~\ref{fig:boxtri}. This integral has six propagators which may be parametrized as
\begin{align}
P_1 &= k_a^2 \,, & P_2 &= (k_a-p_1)^2 \,, & P_3 &= (k_a-p_1-p_2)^2 \,, \nonumber \\
P_4 &= (k_b+p_4)^2 \,, & P_5 &= k_b^2 \,, & P_6 &= (k_a-k_b)^2 \,.
\label{eq:boxtri}
\end{align}
We have named the loop momenta $k_a$ and $k_b$, as opposed to the more standard $k_1$ and $k_2$, in an attempt to make discussion about different orderings of the loops less confusing.
The kinematics is such that
\begin{align}
p_1^2 = p_2^2 = p_3^2 = p_4^2 = 0 \,, \quad (p_1+p_2)^2 = s \,, \quad (p_1+p_4)^2 = t \,, \quad (p_2+p_4)^2 = u = -s-t \,,
\end{align}
but, as mentioned above, that fact is irrelevant for the following discussion, even though it obviously will affect the expressions for the Baikov polynomials in the end.

Before discussing its loop-by-loop parametrizations, let us discuss how to parametrize this integral using the standard Baikov representation. The integral has three independent external momenta $\{p_1,p_2,p_4\}$ and two loops, meaning that $n_{\text{std}}$ as given by eq.~\eqref{eq:nstd} is $9$. This means that we would need three extra propagators to complement the set of eq.~\eqref{eq:boxtri} in order to make the parametrization. If we multiply out the expressions in eq.~\eqref{eq:boxtri} we see that the following six scalar products involving the loop momenta are present:
\begin{align}
\{ k_a^2 \,,\;\; k_b^2 \,,\;\; k_a {\cdot} k_b \,,\;\; k_a {\cdot} p_1 \,,\;\; k_a {\cdot} p_2 \,,\;\; k_b {\cdot} p_4 \}
\label{eq:btdpnat}
\end{align}
It is easy to realize that the missing three are
\begin{align}
\{ k_a {\cdot} p_4 \,,\;\; k_b {\cdot} p_1 \,,\;\; k_b {\cdot} p_2 \}
\label{eq:dpextra}
\end{align}
and one option, in some sense the simplest, for the choice of extra propagators is to directly promote these three scalar products, i.e.
\begin{align}
P_7 &= k_a \cdot p_4 & P_8 &= k_b \cdot p_1 & P_9 &= k_b \cdot p_2
\end{align}
Yet it is often desirable to choose the extra propagators as objects that to a higher degree resemble the propagators already occurring in the problem. Doing so will for instance make certain symmetries of the problem more transparent, and often for phenomenological applications other Feynman integrals that might contribute to the process will have as propagators a different subset of such a more realistic choice. One good choice is
\begin{align}
P_7 &= (k_a + p_4)^2 & P_8 &= (k_b - p_1)^2 & P_9 &= (k_b - p_1 - p_2)^2
\label{eq:dpprextra}
\end{align}
which we see to contain the scalar products of eq.~\eqref{eq:dpextra} upon expanding, and furthermore we see that the whole set of nine propagators will go to itself under the symmetry $k_a \leftrightarrow k_b$. The reader may check that the transformation from these nine propagators to the nine scalar products indeed is invertible.

Needing three extra propagators to make a parametrization seems a bit much, so let us see if we can do better with a loop-by-loop parametrization. 

When making such a parametrization, we have two options. Either we start with the $k_a$-loop or with the $k_b$-loop. Let us first look as the option of starting with $k_a$: The $k_a$-loop has three independent external momenta, which are $\{p_1, p_2, k_b\}$. In other words $E_1=3$. The $\mathcal{E}$ of eq.~\eqref{eq:oneloop} is thus $\det(G(p_1, p_2, k_b))$ which we realize to be given in terms of the scalar products $\{k_b^2, k_b {\cdot} p_1 , k_b {\cdot} p_2\}$. Of those $k_b^2$ is present in the list of eq.~\eqref{eq:btdpnat} but the last two $\{k_b {\cdot} p_1 , k_b {\cdot} p_2\}$ are not. We are forced to introduce them as extra scalar products in the problem.
Continuing with the $k_b$ loop we now see that of the scalar products involving $k_b$ we now have the full set $\{k_b^2, k_b {\cdot} p_1 , k_b {\cdot} p_2 , k_b {\cdot} p_4\}$ meaning that $E_2 = 3$. From eq.~\eqref{eq:nlbl} we get $n_{\text{lbl}} = 8$, one less than before. Thus we need only two extra Baikov variables in this case. One option is to promote the two new scalar products $\{k_b {\cdot} p_1 , k_b {\cdot} p_2\}$ to Baikov variables directly, as in the problem above. However, also as before, it is often desirable to pick some that resemble the propagators already in the problem, and we may thus recycle the choice from eq.~\eqref{eq:dpprextra} and pick
\begin{align}
P_7 &= (k_b - p_1)^2 & P_8 &= (k_b - p_1 - p_2)^2
\label{eq:bt1}
\end{align}
as our two extra Baikov variables. Thus this loop-by-loop parametrization allowed us to only introduce two extra variables as opposed to the three from the standard choice.

Let us however look at the other loop-order and consider the case where $k_1 = k_b$. For the $k_b$-loop there are two independent external momenta $\{p_4, k_a\}$ so $E_1 = 2$. This means that the $\mathcal{E}$ of eq.~\eqref{eq:oneloop} is $\det(G(p_4, k_a))$ and thus given in terms of $k_a^2$ and $k_a \cdot p_4$. Of these $k_a^2$ is already present in the problem, but $k_a \cdot p_4$ is not, forcing us to introduce it as an extra scalar product. Moving on to the $k_a$-loop we now have the full set of scalar products $\{k_a^2, k_a {\cdot} p_1 , k_a {\cdot} p_2 , k_a {\cdot} p_4\}$ present in the problem giving $E_2 = 3$. (Unless a Feynman integral factorizes, it is always the case that $E_L = E$.) Thus we get $n_{\text{lbl}}=7$ for this loop order, meaning that only one extra Baikov variable is needed. We may pick
\begin{align}
P_7 &= (k_a + p_4)^2
\label{eq:bt2}
\end{align}
We therefore see that starting with the $k_b$-loop gives the optimal loop-by-loop parametrization for this diagram.

\subsection{The graphical approach}
\label{sec:graphical}

\begin{figure}
\centering
\vspace{-2mm}
\includegraphics[width=0.60\textwidth]{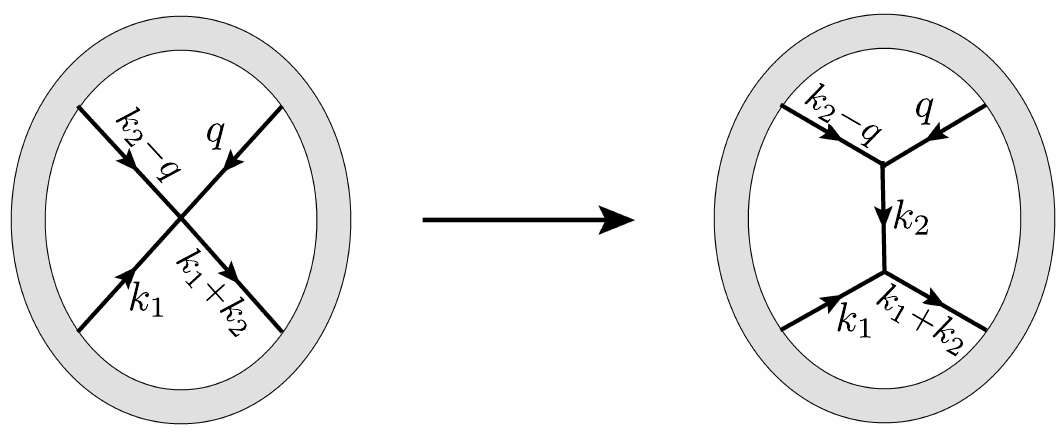}
\vspace{-2mm}
\caption{An example of \textit{opening up} a four-point vertex. The gray blob is meant to indicate that the vertex is embedded in a larger diagram. After integrating out the $k_1$-loop, $\mathcal{E}$ and the remaining propagators will contain the dot-products $\{k_2^2, q^2, k_2 {\cdot} q \}$. After the opening shown, the relation between these dot-products and the remaining propagators will be invertable.}
\vspace{0mm}
\label{fig:openup}
\end{figure}

Baikov parametrizing a given loop will introduce scalar products between all the momenta external to that loop. If the loop is an $n$-gon, it will have $n-1$ independent external momenta, and between them we can form $n(n-1)/2$ such scalar products. Yet (for $n>2$) $n$ of those are already present in the momenta external to the loop, if those momenta correspond to propagators. If not, they will be the first extra propagators to be introduced. This corresponds to \textit{opening up} the vertices of the loop into three-point vertices, in such a way that no further propagators are introduced internally in the loop. This opening up procedure is illustrated in fig.~\ref{fig:openup}. After this is done we are left with $n(n-1)/2 - n$ corresponding to $n(n-3)/2$ different dot-products between the momenta external to the loop, that in principle will have to be mapped to extra propagators. However, if the loop has more than one momentum external to the whole diagram entering it, the scalar products between those will not contain any loop momenta and as such do not need to be mapped to a propagator. Let us call the number of the momenta external to the loop that are also external to the whole diagram $n_e$ (we do not account for one-loop or factorized diagrams here, so we assume that $n_e \leq n-2$). We then get that after making all the vertices of the loop into three-point vertices, there are
\begin{align}
\frac{n (n-3)}{2} - \frac{n_e (n_e-1)}{2}
\label{eq:extracount}
\end{align}
extra scalar products that have to be introduced through extra propagators.

Since our goal is to minimize the number of such extra propagators, we realize that the best strategy here is to start with the loop that has the fewest propagators and then continue in increasing order. This also agrees with the conclusion we drew from the example in sec.~\ref{sec:boxtriangle}. We may in fact split this procedure up into two separate steps, such that in the first step we open up all the higher point vertices, and in the second we introduce the remaining propagators counted by eq.~\eqref{eq:extracount} one loop at a time.\\

Specifically, we now have the two-step prescription:
\begin{itemize}
  \item[Step 1:] Open all higher-point vertices into three-point vertices, in such a way that the number of edges in the smallest of the new one-loop polygons gets minimized.
  \item[Step 2:] Starting from the smallest polygons, parameterize the loops one by one. When new dot-products appear, introduce extra propagators containing them.
\end{itemize}
There is a convenient graphical way to visualize and apply this procedure, as we shall see, and for that reason we name it ``the graphical approach''.

Let us illustrate this approach with a number of examples. We begin with the box-triangle discussed in sec.~\ref{sec:boxtriangle}. That diagram has one four-point vertex and all remaining vertices being three-point. There are two options for opening up that vertex. The one we prefer is the one after which the smallest polygon, the triangle, will remain a triangle. (For the other opening\footnote{That other opening will correspond to introducing the two extra propagators given by eq.~\eqref{eq:bt1}. Step 1 would introduce the extra propagator $P_8$ of eq.~\eqref{eq:bt1}, and step 2, integrating out the $k_a$-loop, would further introduce $P_7$ of eq.~\eqref{eq:bt1}.} we would get two boxes and would be forced to introduce one extra propagator after step 2). This opening that preserves the triangle will introduce one extra propagator, which with the parametrization given by eq.~\eqref{eq:boxtri} will correspond to the $P_7$ given by eq.~\eqref{eq:bt2}.
\begin{align}
\adjustbox{raise=-9mm}{\includegraphics[width=10cm]{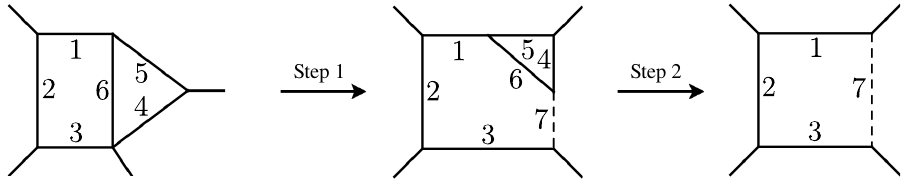}}
\label{eq:boxtristep}
\end{align}
where we have drawn the extra propagator as a dotted line, a notation we will use in the rest of this section. We then get to step 2, where we have to Baikov-parametrize the loops one at a time, starting from the smallest. The scalar product between the two independent momenta external to the triangle subloop is already present in the third of the propagators, so no new dot-products will have to be introduced in this step, as will always be the case for triangle sub-loops after step 1. Then the remaining loop may be Baikov-parametrized as well, leading to the completed loop-by-loop parametrization.

Next we will have a look at the three-loop banana integral. It has two five-point vertices that can be opened as shown under step 1:
\begin{align}
\adjustbox{raise=-5mm}{\includegraphics[width=14cm]{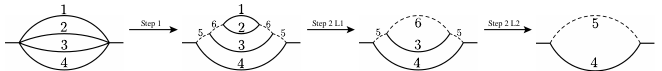}}
\end{align}
Note that for a bubble (or in general any two-point subdiagram) the propagator on each side can be chosen to be the same, so only two new propagators (5 and 6) are introduced in this step. This is our only example where the counting of eq.~\eqref{eq:extracount} fails.
We will in general not discuss specific parametrizations in this section, since the approach discussed here is parametrization independent. However, if the banana initially is parametrized as
\begin{align}
P_1 = k_a^2 \,,\quad P_2 = (k_a-k_b)^2 \,,\quad P_3 = (k_b-k_c)^2 \,,\quad  P_4 = (k_c-p)^2 \,,
\end{align}
the two extra propagators will be
\begin{align}
P_5 = k_c^2 \,,\quad P_6 = k_b^2 \,.
\end{align}
Then we may Baikov parametrize the bubble with propagators 1 and 2 without introducing any new propagators, and then the same with the bubble with propagators 3 and 6, and lastly the remaining bubble finishing the parametrization.

Our next example will be the two-loop crossed box. It only has three-point vertices so step 1 is trivial. We then have to choose which loop to start with, and we pick the crossed side of the diagram, as that loop has four propagators whereas the other one has five. Baikov-parametrizing that four-propagator loop, will introduce three scalar products between the three independent momenta external to the loop. One of them will correspond to the already present fourth leg of that loop. Another is between two momenta external to the integral, and will therefore not contain any loop momenta and not correspond to a propagator. That leaves one, as given by eq.~\eqref{eq:extracount}, which will correspond to a new propagator that will have to be introduced as a $P_8$ as shown:
\begin{align}
\adjustbox{raise=-6mm}{\includegraphics[width=5cm]{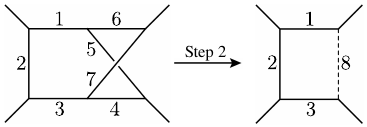}}
\end{align}
This leaves the remaining loop, which now can be parametrized as well. Let us mention that the strategy for the planar doublebox would be exactly similar.

It sometimes happens that the graphical approach of this section is not as ``linear'' as in the previous examples. That happens when more than one extra propagator has to be introduced after parametrizing a loop in step 2, in such a way no diagram containing them all can be (conveniently) drawn. In that case the intermediate stages of the approach will have to contain a ``superposition'' of diagrams, which together contain the complete set of propagators. This happens whenever three (or more) loops that are boxes or above, all touch each other. A paradigmatic example of this is the tennis court, which we will consider next:
\begin{align}
\adjustbox{raise=-19mm}{\includegraphics[width=8cm]{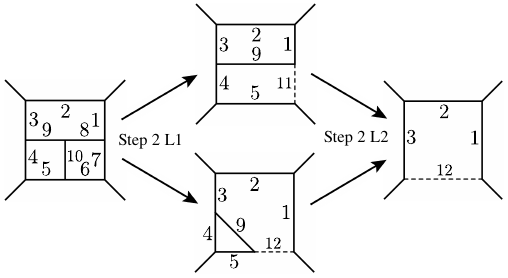}}
\end{align}
This diagram has only three-point vertices so step 1 is trivial. As the first loop to parametrize in step 2 we must pick one of the two boxes, and we pick the one in the lower right corner with propagators $\{6,7,8,10\}$. Three scalar products can be formed between the three independent legs of that box. One is contained in the fourth leg, but the remaining two have to be introduced as extra propagators $P_{11}$ and $P_{12}$. No diagram that contains the remaining six of the original propagators as well as $P_{11}$ and $P_{12}$, but no more, can be conveniently\footnote{It is possible to draw it as a multiply collinear limit of a six-point two-loop diagram. Doing so will, however, confuse more than it clarifies. This ``superposition'' approach is much preferable.} drawn. For that reason we have to resort to the ``superposition'' approach for the depiction of that step. Next we will parametrize the loop that contains the lower left corner, since that loop has four propagators $\{4,5,9,11\}$ whereas the other has five $\{1,2,3,9,12\}$. Parametrizing that loop introduces three scalar products, one of which is between two external momenta, and the last two are represented in $P_3$ and $P_{12}$ so no new propagators are needed. We may then parametrize the last loop, finishing the job.

Next we will consider a non-planar three-loop four-point diagram\footnote{Thanks to William J. Torres Bobadilla for suggesting me this example.}. It is symmetric in its three loops; we will begin by Baikov-parametrizing the one shown on the right, i.e. the one with propagators numbered $\{5,6,7,8,9\}$. Following eq.~\eqref{eq:extracount}, six scalar products can be formed between its four independent legs. One corresponds to the fifth leg, one is between two external momenta, but the last four will have to be introduced as new propagators $P_{11}$--$P_{14}$. We could show these four in a ``superposition'' of two figures (one with $P_{11}$ and $P_{12}$ and the other with $P_{13}$ and $P_{14}$), but find it more clarifying to introduce a third figure bridging the two:
\begin{align}
\adjustbox{raise=-26mm}{\includegraphics[width=8cm]{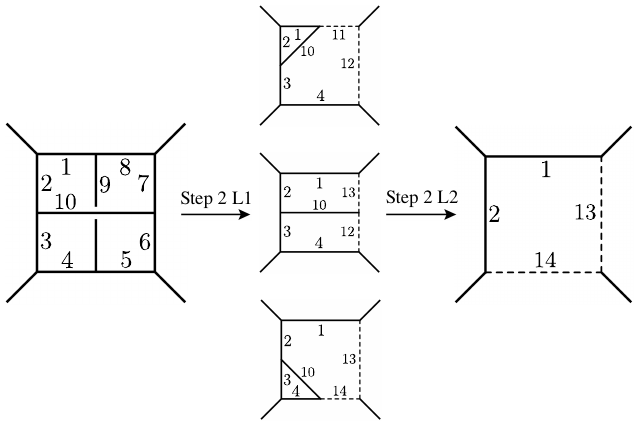}}
\end{align}
We may now parametrize the second loop, picking the lower one with propagators $\{3,4,10,11,12\}$. Of the six scalar products thereby introduced, three are between external momenta, and the remaining three correspond to propagators already in the problem, so no new propagators have to be introduced in this step. Lastly we may parametrize the third loop, finishing the exercise.

As a last example we will look at a four-loop four point integral. The parametrization of that integral will require ``superpositions'' at two of the intermediate steps, but other than that it proceeds as described above, and we will present it without further comments:

\begin{align}
\adjustbox{raise=-30mm}{\includegraphics[width=13cm]{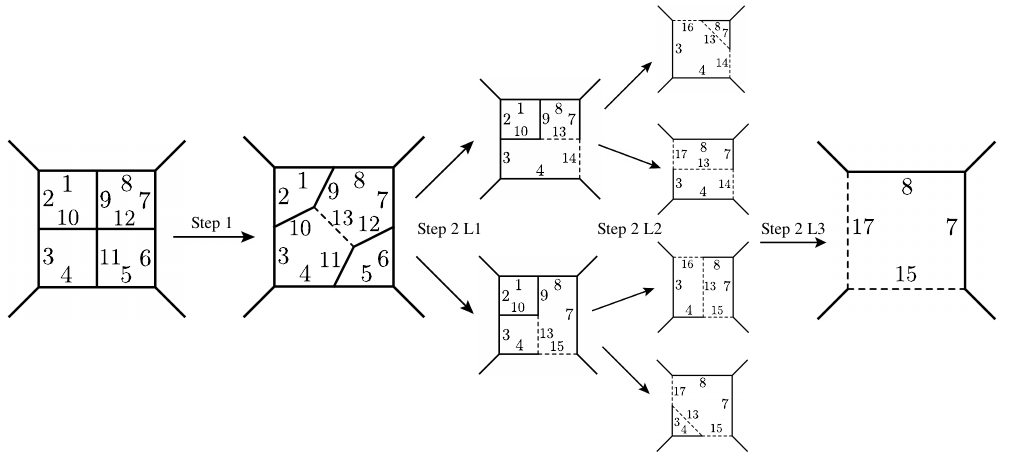}}
\end{align}

We hope with this discussion to have given the reader the tools to make a good\footnote{It should be noted that there are cases~\cite{Brammer:paper} in which a loop-by-loop parametrization with a number of variables that is \textit{higher} than that following from the above algorithm, is desirable for a specific purpose such as extracting a leading singularity. We expect such cases to be exceedingly rare.} loop-by-loop Baikov parametrization of any Feynman integral that may be encountered.

\section{The Baikov Package}
\label{sec:BaikovPackage}

In this section we will discuss my implementation of the Baikov representation in its standard and loop-by-loop varieties. The implementation is made as a \texttt{Mathematica} package named \texttt{BaikovPackage} which can be downloaded from \url{https://github.com/HjalteFrellesvig/BaikovPackage}. At that link can be found a text-file \texttt{BaikovPackage.m} containing the package itself, along with a \texttt{Mathematica} notebook \texttt{BaikovPackageExamples.nb} containing examples of its use. Further examples may be seen in the figure on page \pageref{fig:screenshot}. Please note the absence of dependencies, and the fact that no installation is needed, the package may be loaded with \texttt{Mathematica}'s \texttt{Get} command. In the following we will describe the use and the various features of \texttt{BaikovPackage}.

\subsection{Setting up the problem}
\label{sec:settingup}

To set up a specific Feynman integral family, one needs to assign values to five pre-reserved objects called \texttt{Internal}, \texttt{External}, \texttt{Propagators}, \texttt{PropagatorsExtra}, and \texttt{Replacements}.\\

\texttt{Internal} is a list containing the independent internal momenta (or loop momenta) of the problem. Thus for an $L$-loop Feynman integral, \texttt{Internal} should have $L$ elements. In examples we will tend to name the members of \texttt{Internal} from $k_1$ to $k_L$, but no specific naming convention is required for \texttt{BaikovPackage} to function. For the loop-by-loop Baikov representation the loop ordering in encoded in \texttt{Internal}, in that the loops get parametrized in the order from \textbf{right to left} in that list. For the standard Baikov representation the ordering of \texttt{Internal} is irrelevant.\\

\texttt{External} is a list containing the independent external momenta of the problem. An integral with $E$ such independent external momenta will usually have $E+1$ external legs. In examples we will tend to name the members of \texttt{External} from $p_1$ to $p_E$, but also here no specific naming convention is required for \texttt{BaikovPackage} to function.\\

\texttt{Propagators} is the list of (actual, physical, and linearly independent\footnote{If the set of propagators is not linearly independent \texttt{BaikovPackage} cannot be applied directly. What to do in such cases is discussed in app.~\ref{app:cannot}.}) propagators present in the Feynman integral. These propagators must be described in terms of the elements of \texttt{Internal} and \texttt{External}, as well as potential internal masses for which (again) no specific naming convention is required. The propagators can be quadratic in the loop momenta as it usually is the case in quantum field theory, but linear propagators (of the type that appear in soft expansions and in the post-Minkowskian approach to classical gravity) are valid as well.\\

\begin{figure}
\centering
\vspace{-2mm}
\includegraphics[width=0.65\textwidth]{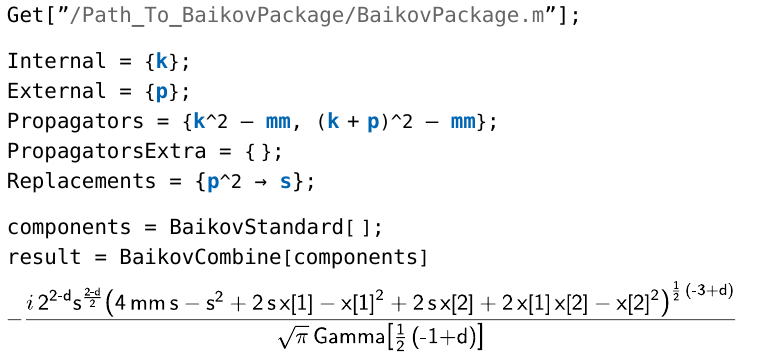}
\vspace{-1mm}
\caption{A simple example of the use of \texttt{BaikovPackage} to generate the Baikov representation for a one-loop bubble integral.}
\vspace{-1mm}
\label{fig:smallcodeexample}
\end{figure}

\texttt{PropagatorsExtra} is the list of extra propagator-type objects that have to be introduced in order to achieve a valid Baikov parametrization\footnote{It should be noted that the separation into \texttt{Propagators} and \texttt{PropagatorsExtra} is purely psychological, in that one of the first lines of code is \texttt{AllPropagators = Join[Propagators, PropagatorsExtra];} and then the two are not mentioned again. Yet the psychological effect of the distinction is so strong that it is worth keeping, since the members of \texttt{Propagators} are dictated (up to parametrization and ordering) by the physical problem, whereas the expressions and even the number of the elements of \texttt{PropagatorsExtra} can be changed and optimized by the user.}. This list should have $N_{\text{extra}} = n-P$ members where $n$ is given by eq.~\eqref{eq:nstd} for the standard Baikov representation, and by eq.~\eqref{eq:nlbl} for the loop-by-loop Baikov representation. How to find $N_{\text{extra}}$ as well as expressions for the elements of \texttt{PropagatorsExtra} in the case of loop-by-loop was the topic of sec. \ref{sec:strategies}. Let me clarify that (at the moment) \texttt{BaikovPackage} cannot help with those tasks.\\

\texttt{Replacements} is the list of \texttt{Mathematica} replacement rules that replace (scalar) products of the independent external momenta of \texttt{External} with kinematic variables, be they Mandelstam variables or external masses. It is important that all combinations of external momenta get replaced, so \texttt{Replacements} should have $E(E{+}1)/2$ elements.\\

It may be noted that these conventions and notations are very similar to those used by the IBP program \texttt{FIRE}~\cite{Smirnov:2019qkx}, and that is indeed where the inspiration comes from. The only two differences are the fact that the ordering of the members of \texttt{Internal} matters for the loop-by-loop parametrization, which it never does in \texttt{FIRE}, as well as the distinction between \texttt{Propagators} and \texttt{PropagatorsExtra} which is not present in \texttt{FIRE}. \texttt{FIRE} has no \texttt{PropagatorsExtra} and puts them all into \texttt{Propagators}.

\subsection{Obtaining a Baikov representation}

We will now present the three functions \texttt{BaikovStandard}, \texttt{BaikovLBL}, and \texttt{BaikovCombine}, which are used to obtain the desired Baikov representation. Please see fig.~\ref{fig:smallcodeexample} for a basic illustration of their use.

The components of the Baikov representation get returned by the function \texttt{BaikovStandard} for the case of the standard Baikov representation, and by the function \texttt{BaikovLBL} for the case of loop-by-loop. We will discuss the exact nature of those components in the following section; the focus here is on how to use them. \texttt{BaikovStandard} and \texttt{BaikovLBL} take no arguments, since the variables defining the Feynman integral family (i.e. the five objects discussed in sec.~\ref{sec:settingup}) get passed to \texttt{BaikovStandard} and \texttt{BaikovLBL} as global variables. Those returned components get assembled into the final expression by a third function, \texttt{BaikovCombine}, which takes the output of \texttt{BaikovStandard} or \texttt{BaikovLBL} as input, and returns the expression given by eq.~\eqref{eq:standard} for the case of standard Baikov, and the expression given by eq.~\eqref{eq:loop-by-loop} for the case of loop-by-loop.

The output will be expressed in terms of $n$ Baikov variables, which are named \texttt{x[}$i$\texttt{]} where $i$ goes from $1$ to $n$. Of those variables \texttt{x[}$1$\texttt{]} to \texttt{x[}$P$\texttt{]} correspond to the members of \texttt{Propagators} (in the same order) while \texttt{x[}$P{+}1$\texttt{]} to \texttt{x[}$n$\texttt{]} correspond to the members of \texttt{PropagatorsExtra} (also in the same order). Furthermore there will be dependence on the various kinematic variables and on the spacetime dimensionality \texttt{d}. Note that both \texttt{x} and \texttt{d} are reserved variable names in \texttt{BaikovPackage}.

Compared to eqs.~\eqref{eq:standard} and \eqref{eq:loop-by-loop}, the propagators themselves, i.e. the factors of $x_i^{-a_i}$, will not be included in the output of \texttt{BaikovCombine}. That is done in order to avoid passing the values of the $a_i$ to the various functions, but it is of course trivial for the user to include these factors afterwards. Furthermore no information on the integration region $\mathcal{C}$ is provided by the functions in \texttt{BaikovPackage}, only the integrand is returned by \texttt{BaikovCombine}.

\subsection{The Baikov component format}

It is worth having a closer look at the output format of \texttt{BaikovStandard} and \texttt{BaikovLBL}. Let us save the output of one of these functions as a \texttt{Mathematica} object named \texttt{Comp} (not a reserved name in \texttt{BaikovPackage}). \texttt{Comp} will be a list with four elements, that we will now discuss.

\texttt{Comp[[1]]} is a list containing the Baikov polynomials. For the two cases of standard and loop-by-loop Baikov this list is
\begin{align}
\texttt{Comp[[1]]}_{\text{std}} = \left\{ \mathcal{E}, \mathcal{B} \right\} \; , \qquad \texttt{Comp[[1]]}_{\text{lbl}} = \left\{ \mathcal{E}_L, \mathcal{B}_L , \ldots , \mathcal{E}_1, \mathcal{B}_1 \right\}
\end{align}
and thus is has length $2$ for standard Baikov, and length $2L$ for loop-by-loop. For loop-by-loop the ordering is such that the subscript on the $\mathcal{E}_i$ and $\mathcal{B}_i$ refers to the $i$th loop to be parametrized and thus the ordering is the same as in \texttt{Internal}.

\texttt{Comp[[2]]} is a list with same length as \texttt{Comp[[1]]}, containing the powers to which those polynomials are raised in eqs.~\eqref{eq:standard} and \eqref{eq:loop-by-loop}, that is
\begin{align}
\texttt{Comp[[2]]}_{\text{std}} &= \left\{ \frac{E{-}d{+}1}{2} \,,\; \frac{d{-}E{-}L{-}1}{2} \right\} \nonumber \\
\texttt{Comp[[2]]}_{\text{lbl}} &= \left\{ \frac{E_L{-}d{+}1}{2} \,,\; \frac{d{-}E_L{-}2}{2} \,,\; \ldots \,,\; \frac{E_1{-}d{+}1}{2} \,,\; \frac{d{-}E_1{-}2}{2} \right\}
\end{align}

\texttt{Comp[[3]]} contains the prefactors of the integral in the following format:
\begin{align}
\texttt{Comp[[3]]}_{\text{std}} &= \left\{ \frac{(-i)^L \, \pi^{(L-n)/2}}{\prod_{l=1}^L \Gamma \left( (d{+}1{-}E{-}l)/2 \right)} \,,\; \mathcal{J} \right\}  \nonumber \\
\texttt{Comp[[3]]}_{\text{lbl}} &= \left\{ \frac{-i \, \pi^{-E_L/2}}{\Gamma \left( (d{-}E_L)/2 \right)} \,,\; \cdots \,,\; \frac{-i \, \pi^{-E_1/2}}{\Gamma \left( (d{-}E_1)/2 \right)} \,,\; \mathcal{J} \right\}
\end{align}
and thus it has two members for standard Baikov and $L+1$ members for loop-by-loop. We see that in either case, the constant prefactor of the integral is the product of all the elements of \texttt{Comp[[3]]}.

From these ingredients we are now able to obtain the combined integrand for the Baikov parametrization (excluding the propagators) in either case, by
\begin{align}
\bigg( \prod_j \texttt{Comp[[3}, j \texttt{]]} \bigg) \times \prod_i \texttt{Comp[[1}, i \texttt{]]}^{\texttt{Comp[[2}, i \texttt{]]}}
\end{align}
which is exactly how it is implemented in \texttt{BaikovCombine}.

\texttt{Comp[[4]]} contains information that while not directly needed to build the Baikov representation, could still be useful to the user. \texttt{Comp[[4]]} is itself a list with four elements (a number that could become higher in the future if more features are added).

\texttt{Comp[[4,1]]} contains replacement rules that replace products of momenta, of which one is a member of \texttt{Internal} and the other either of \texttt{Internal} or \texttt{External}, with a \texttt{Mathematica} object called \texttt{DP} (for Dot Product) used to store them in intermediate steps. \texttt{DP} is a reserved name in \texttt{BaikovPackage}. All combinations will be replaced regardless of whether they appear in a loop-by-loop parametrization, so \texttt{Comp[[4,1]]} will have $L(L{+}1)/2 + LE$ elements.

\texttt{Comp[[4,2]]} contains replacement rules to go from the dot products in the \texttt{DP} format to the Baikov variables in the \texttt{x} format. There will be as many elements here as there are Baikov variables, i.e. $n$, so in standard Baikov all possible \texttt{DP}s will be replaced, whereas that is usually not the case for loop-by-loop.

\texttt{Comp[[4,3]]}: For the loop-by-loop representation, \texttt{Comp[[4,3]]} contains the momenta considered external to the individual loops. So for an $L$-loop problem, \texttt{Comp[[4,3]]} will have $L$ elements, each of which will be a list of momenta, such that the $i$th element from the right will be the list of momenta external to the $i$th loop in the integration order. A list of $E_i$ can be obtained by mapping \texttt{Length} to \texttt{Comp[[4,3]]}. For standard Baikov \texttt{Comp[[4,3]]} will have one element which is set to \texttt{External}.

\texttt{Comp[[4,4]]} is related to the scalar products appearing internally, and will be discussed shortly.

\begin{figure}
\centering
\vspace{-2mm}
\includegraphics[width=0.99\textwidth]{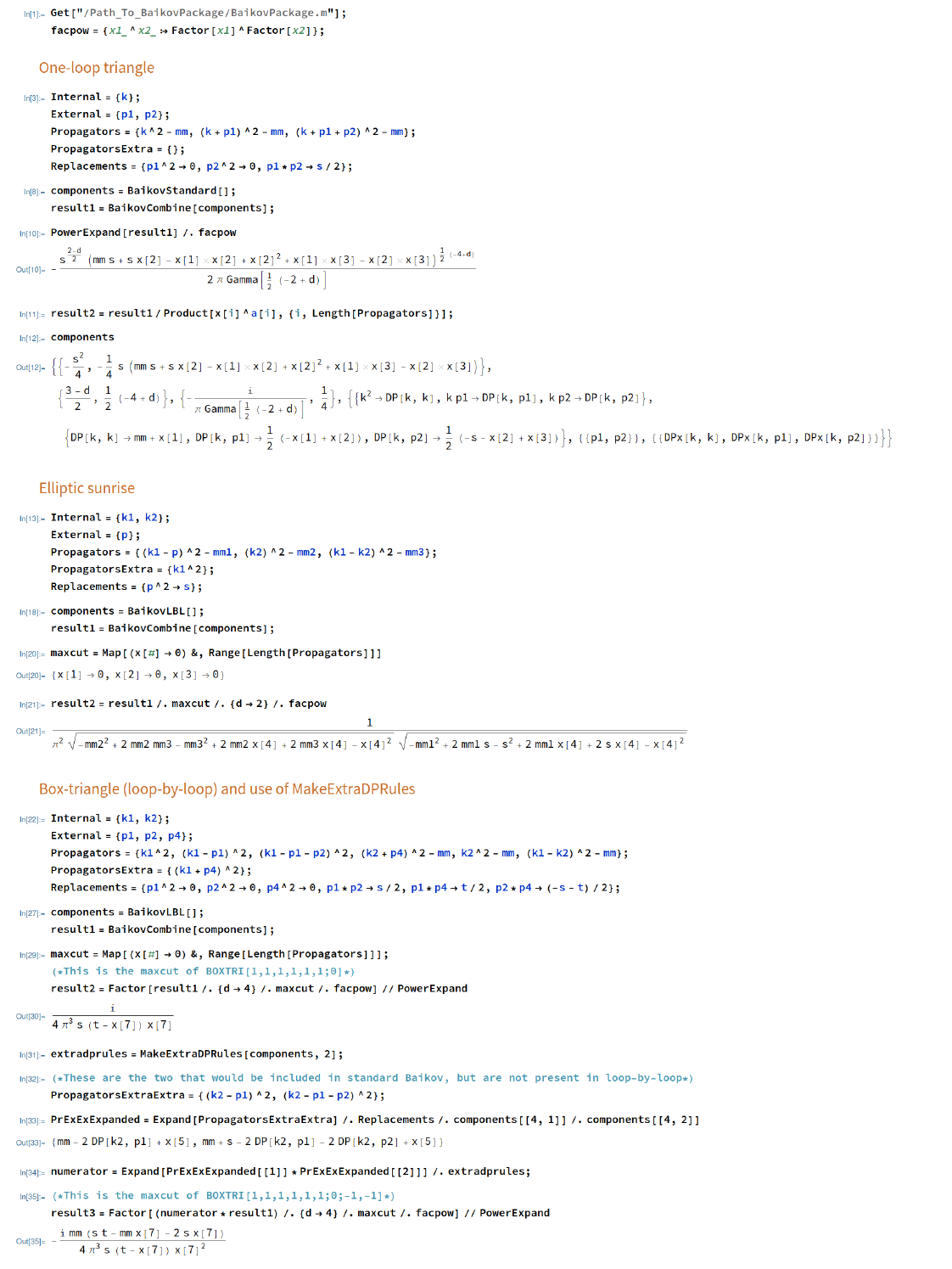}
\vspace{-2mm}
\caption{Screenshot with three examples of the use of \texttt{BaikovPackage}.}
\vspace{-4mm}
\label{fig:screenshot}
\end{figure}

\subsection{Additional features and settings}
\label{sec:features}

We will now discuss a number of additional\footnote{Warning: The functionalities discussed in this section, in particular \texttt{MakeExtraDPRules} and \texttt{BPDPresult}, are less debugged and stress tested than the rest of \texttt{BaikovPackage}.} features and settings of \texttt{BaikovPackage}.

\texttt{MakeExtraDPRules} is an additional function in \texttt{BaikovPackage}. This function generates rules that may be applied to scalar products that are not included in a loop-by-loop parametrization, in order to make that loop-by-loop parametrization applicable to cases where such scalar products appear in the numerator. This function works as described in appendix \ref{app:remaining}. The function is called as \texttt{MakeExtraDPRules[Comp, rank]}, where \texttt{Comp} is the output of \texttt{BaikovLBL} and \texttt{rank} ($\tau$ in app.~\ref{app:remaining}) is a positive integer denoting the highest power of these extra scalar products\footnote{If we call the dot-products mappable to members of \texttt{Propagators} RSPs (reducible scalar products) and dot-products mappable to members of \texttt{PropagatorsExtra} ISPs (irreducible scalar products), then those in this remaining set will be XSPs (extra scalar products). There will be $P$ RSPs, $(n_{\text{lbl}}{-}P)$ ISPs, and $(n_{\text{std}}{-}n_{\text{lbl}})$ XSPs, corresponding to $n_{\text{std}} = L(L{+}1)/2 + LE$ scalar products in total.} for which we are interested in such replacement rules. This \texttt{rank} is needed since its value affects the runtime of \texttt{MakeExtraDPRules} very significantly as discussed in app.~\ref{app:remaining}. When a Feynman integral has a numerator $N(k)$ expressed in terms of scalar products between momenta, these should be expressed in terms of \texttt{DP} objects. Those that can be mapped to the Baikov variables can be handled with \texttt{Comp[[4,2]]}, and the remaining ones can be handled with the output of \texttt{MakeExtraDPRules} after expanding, yielding an expression $N(x)$ given solely in terms of the variables of the loop-by-loop representation.\\

The functions in \texttt{BaikovPackage} come with a number (currently three) of \textit{settings} that can be turned on or off. Those settings are stored in the internal control variables \texttt{BPprint}, \texttt{BPFactorFinal}, and \texttt{BPDPresult}. These variables are not directly accessible to the user, but they can be manipulated using \textit{get} and \textit{set} functions. The get-functions \texttt{GetBPprint}, \texttt{GetFactorFinal}, and \texttt{GetDPresult} take no arguments, and return the value of the corresponding control variable. The set-functions \texttt{SetBPprint}, \texttt{SetFactorFinal}, and \texttt{SetDPresult}, can be called with either \texttt{True} or \texttt{False} and they will then set the corresponding control variable to that value.\\

\texttt{BPprint} (which is false by default) will, if true, make \texttt{BaikovLBL} and \texttt{MakeExtraDPRules} print status messages at intermediate steps. These messages are for debugging purposes and will probably not be useful for most users.\\

\texttt{BPFactorFinal} (which is true by default) will, if true, apply \texttt{Factor} to the Baikov polynomials in \texttt{Comp[[1]]} before returning from \texttt{BaikovStandard} or \texttt{BaikovLBL}. We allow this factoring to be turned off, since it can take some time for large expressions, particularly when the kinematics of the problem is five-point or above.\\

\texttt{BPDPresult} (which is false by default) will, if true, make the functions \texttt{BaikovStandard} and \texttt{BaikovLBL} return expressions in terms of a set of dot-products involving the loop momenta\footnote{Thanks to Pierpaolo Mastrolia for encouraging me to implement this feature.}, rather than in terms of the Baikov variables \texttt{x}. These dot-product are denoted \texttt{DPx} and they are similar to the \texttt{DP}, except that they are between the loop momenta in \texttt{Internal} and the momenta external to each loop (i.e. those that are returned as \texttt{Comp[[4,3]]}). This corresponds to the representation given by eq.~\eqref{eq:dotprodrep} which is the second-to-last step in the derivation of the Baikov representation, with the \texttt{DPx}-variables corresponding to the $\varsigma_i$-variables found there. \texttt{Comp[[4,4]]} will as its first element have the list of these \texttt{DPx}-variables, and when \texttt{BPDPresult} is true \texttt{Comp[[4,4]]} will as a second element have a list of rules for how to go from the propagators, denoted \texttt{x}, to the \texttt{DPx} variables. Furthermore \texttt{Comp[[4,2]]} will become rules for going from \texttt{DP} to \texttt{DPx}. Also when \texttt{BPDPresult} is true the Jacobian $\mathcal{J}$, which is the last entry in \texttt{Comp[[3]]}, will equal one.

\section{Perspectives}
\label{sec:perspectives}

The Baikov representation is significantly less studied than the Feynman parameter representation and its related representations. As such there are a number of open problems that we will now discuss.

\subsection{Momentum parametrization invariance}

It is worthwhile discussing the degree to which the Baikov representation, as well as the various functions in \texttt{BaikovPackage}, depend on the initial momentum parametrization of the diagram.

For the result of a Baikov parametrization, either the standard as given by eq.~\eqref{eq:standard} or the loop-by-loop as given by eq.~\eqref{eq:loop-by-loop}, that discussion is over quickly since those results have no dependence on the momentum parametrization. Had it been otherwise that fundamental reparametrization symmetry of Feynman integrals would be violated and the Baikov representation would be invalid. 

There is of course a dependence on the choice of which extra propagator-like objects to promote to Baikov variables in cases where the original set is not large enough. And while that choice is often discussed in terms of a specific momentum parametrization, once the choice has been made a change in the momentum parametrization will change nothing in the final result.

\begin{figure}
\centering
\includegraphics[width=12.5cm]{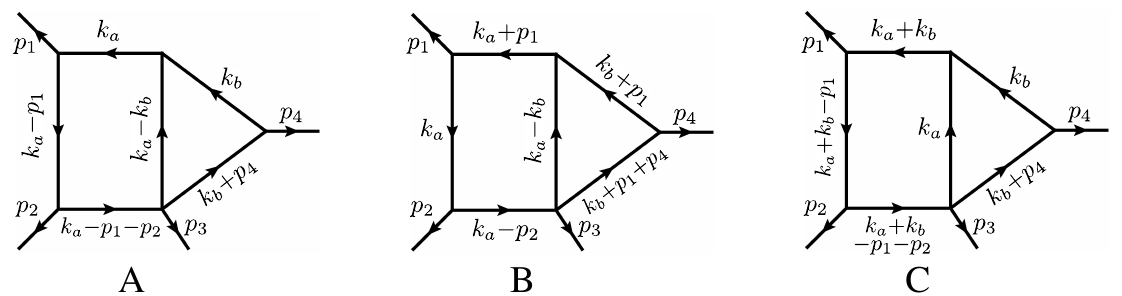}
\vspace{-1mm}
\caption{Three momentum parametrizations of the box-triangle integral.}
\label{fig:parametrizations}
\end{figure}

The functions in \texttt{BaikovPackage} also have a large degree of momentum parametrization independence but it is not complete. We will discuss this in terms of the three parametrizations of the box-triangle shown in fig.~\ref{fig:parametrizations}. The momentum parametrization of fig.~\ref{fig:parametrizations}A is the same as that shown in fig.~\ref{fig:boxtri} and discussed in sec.~\ref{sec:boxtriangle}. We see explicitly when looking at the triangle subloop, that in addition to the loop momentum $k_b$ only $k_a$ and $p_4$ appears, so it is clear that $E_b=2$ and what those two are.

\begin{figure}
\centering
\includegraphics[width=5.5cm]{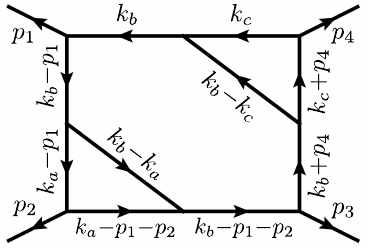}
\vspace{0mm}
\caption{A three-loop integral that requires high momentum parametrization independence to Baikov parametrize well. \texttt{BaikovPackage} is up to the task.}
\label{fig:threeloop}
\end{figure}

If we, however, do the reparametrization $k_i \rightarrow k_i + p_1$ we obtain the momentum parametrization of fig.~\ref{fig:parametrizations}B. Looking again at the triangle sub-loop, we now in addition to the loop momentum $k_b$ have dependence on $k_a$, $p_1$, and $p_4$, so it might appear as if $E_b=3$ forcing us to introduce an extra Baikov variable accounting for this extra scalar product. Yet the implementation in \texttt{BaikovPackage} does not make this mistake. It is able to notice that the reparametrization $k_b \rightarrow k_b-p_1$ would make $k_a$ and $p_1$ appear only through their sum, so that loop is correctly seen to have $E_b=2$, corresponding to the set of externals $\{ p_4, k_a{+}p_1 \}$ (as can be seen in \texttt{Comp[[4,3]]}). \texttt{BaikovPackage} having this ability is convenient, but there are cases where it is not just convenient but necessary\footnote{Thanks to Roger Morales for making me aware of examples such as this.} such as the example shown in fig.~\ref{fig:threeloop}.

As by the discussion in sec.~\ref{sec:graphical}, a loop-by-loop Baikov parametrization of the integral shown in fig.~\ref{fig:threeloop} should start with the two triangles in the corners. The $k_c$ triangle is momentum parametrized such that it is obvious that it has $E_c=2$. The $k_a$ triangle on the other hand has dependence on $p_1$, $p_2$, and $k_b$, so it might at a first glance appear to have $E_a=3$. Only a reparametrization analysis as discussed above reveals it to have $E_a=2$. The point is that there is no momentum parametrization of this integral for which determining the $E_i$ is straight forward for all the loops at once, so being able to perform such reparametrizations ``at runtime'' is necessary in order to be able to make an optimal loop-by-loop Baikov parametrization of this integral.

There is however another type of momentum reparametrizations, exemplified by fig.~\ref{fig:parametrizations}C. Compared to fig.~\ref{fig:parametrizations}A, we have here reparametrized one of the loop momenta with a term that includes the other loop momentum, specifically $k_a \rightarrow k_a + k_b$. After this reparametrization, the $k_a$-loop contains 4 propagators and the $k_b$-loop contains 5, so from the parametrization alone it is not easy to determine that this integral has a triangle subloop. Unfortunately \texttt{BaikovPackage} can not handle such cases, in the sense that it will not be able to produce a Baikov parametrization containing only one extra propagator. I do not believe this to be a problem for realistic use cases, since it is not difficult to avoid such parametrizations \textit{ab initio}. Yet the situation is unsatisfying, and it would be great to implement this functionality in the future.

The source of this problem is that \texttt{BaikovPackage} uses the parametrization to determine which subsets of propagators are considered to belong to the same loop. A general $L$-loop diagram will contain up to $2^L-1$ different \textit{cycles}, each of which could potentially be considered a loop in the context of loop-by-loop parametrization. How to assign this in the optimal way is a question properly belonging to graph theory. For the Feynman parameter representation and its related representations, the two Symanzik polynomials $\mathcal{U}$ and $\mathcal{F}$ (which in those representations play the role that the Baikov polynomials $\mathcal{B}_i$ and $\mathcal{E}_i$ play for the loop-by-loop Baikov representation) can be derived purely using graph theory, invoking notions such as \textit{spanning trees} and \textit{2-forests}~\cite{Bogner:2010kv}. An implementation of the Feynman parameter representation might be set up to be called with a specific momentum parametrization, but then that parametrization is used only to generate the underlying graph, and it is then that graph which is used to generate the $\mathcal{U}$ and $\mathcal{F}$-polynomials and thus the Feynman parametrization itself. Deriving and implementing something similar for the Baikov representation would be a worthwhile endeavour.

\subsection{The role of dimensional regularization and the 4D Baikov representation}
\label{sec:4D}

\begin{figure}
\centering
\includegraphics[width=9cm]{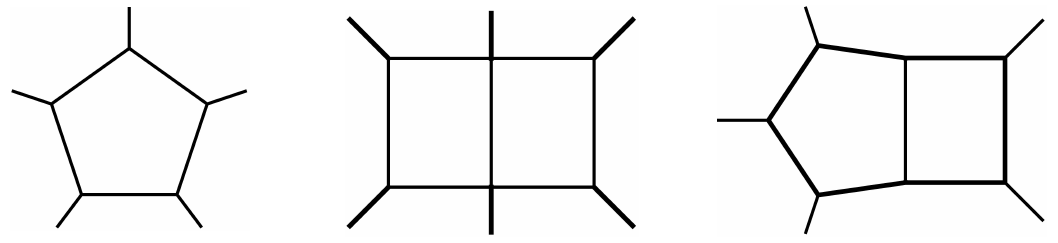}
%\vspace{0mm}
\caption{The one-loop pentagon, the elliptic doublebox, and a finite pentabox.}
\label{fig:pgondbox}
\end{figure}

The Baikov representation is more reliant on dimensional regularization than the Feynman parameter representation and its relatives. This is exemplified by the one-loop pentagon integral, shown in fig.~\ref{fig:pgondbox}, in $4 - 2 \epsilon$ dimensions. In that case eq.~\eqref{eq:standard} gives
\begin{align}
I_{\text{pentagon}} &= \frac{-i \mathcal{J} \, \mathcal{E}^{1/2 + \epsilon}}{\pi^2 \, \Gamma ( - \epsilon )} \int_{\mathcal{C}} \frac{ \mathcal{B}^{-1- \epsilon} \, \id^{5} x }{x_1 \cdots x_5}
\label{eq:pentagon}
\end{align}
If the propagators are massive this pentagon integral is finite. From eq.~\eqref{eq:pentagon} we do, however, see that the prefactor $1/\Gamma(-\epsilon) = - \epsilon + \mathcal{O}(\epsilon^2)$ vanishes in the $\epsilon \rightarrow 0$ limit, whereas the integral part will diverge in that limit since the integration will become of $\mathcal{B}^{-1}$ between zeros of $\mathcal{B}$. The dimensional regularization procedure ensures that the result comes out correctly, but it might still be worrying that dimensional regularization plays such a crucial role for the evaluation of a finite object. The Feynman parameter representation and its related representations will have no such behaviour for this type of integral. This raises the question about whether it is possible to construct an explicitly 4D variant of the Baikov representation, that can be applied to such finite integrals without relying on dimensional regularization.

The origin of this behaviour is the steps in the derivation of the Baikov representation corresponding to eqs.~\eqref{eq:intsep} to \eqref{eq:intstep1}, that separates the loop-integration into parallel and perpendicular components
\begin{align}
\text{d}^d k = \text{d}^E k_{||} \text{d}^{d-E} k_{\perp}
\end{align}
followed by performing the angular part of the perpendicular integral
\begin{align}
\text{d}^{d-E} k_{\perp} \, \rightarrow \; \frac{\pi^{(d-E)/2}}{\Gamma((d-E)/2)} (k^2_{\perp})^{(d-E-2)/2} \, \text{d} (k^2_{\perp})
\end{align}
We see that if from the beginning we have $d=E$, this separation is invalid, we just have $d^d k = d^E k_{||}$ with no perpendicular component introduced. Continuing through the derivation we will then change integration variables from $d^E k_{||}$ to integrations over the dot-products $k \cdot p_1$ to $k \cdot p_4$. $k^2$ will not enter as an independent dot-product at any point.
This four dimensional constraint on $k$ may also be written as the Gram determinant constraint
\begin{align}
\det \! \big( G ( k, p_1, p_2, p_3, p_4 ) \big)  &= 0 
\label{eq:pentagram}
\end{align}
similarly to how the finite dimensionality of spacetime can put constraints on external kinematics as discussed in app.~\ref{app:spacetime}. This constraint reflects the fact that the set of five dot-products, $k^2$ and $k \cdot p_1$ to $k \cdot p_4$, cannot be truly independent in 4D.
We may then write the five dot-products in terms of the five propagators of the pentagon, and isolate the term in eq.~\eqref{eq:pentagram} that is constant in the propagators. This we may then insert in the numerator of the pentagon integral, yielding an expression for the pentagon in terms of sub-sectors valid\footnote{That expression may then be upgraded to an expression valid in $d=4-2 \epsilon$ up to a term of $\mathcal{O}(\epsilon)$ which often can be identified with the pentagon integral in $d = 6 - 2 \epsilon$~\cite{Binoth:1999sp, Bern:1993kr}. Deriving a more explicit expression would bring us too far afield, but see e.g. sec. 5.2 of ref.~\cite{Weinzierl:2022eaz} for further discussion.} in $d=4$. So while the pentagon integral in dimensional regularization is a master integral irreducible to sub-sectors, the truly independent parts of that integral will only enter at $\mathcal{O}(\epsilon)$.

In the loop-by-loop Baikov representation this procedure may also be performed at the level of the individual loops. This is discussed in ref.~\cite{Frellesvig:2021vdl} under the name of the \textit{4D Baikov representation}. The discussion there is done in some generality, but is focused on the elliptic doublebox integral shown in fig.~\ref{fig:pgondbox}. Let us, however, start by considering a different example, that of the pentabox also shown in fig.~\ref{fig:pgondbox}. As with the pentagon, we will consider some distribution of internal and external masses that makes the pentabox finite in $d=4$. If we were to make a loop-by-loop Baikov parametrization of the pentabox using the method discussed in sec.~\ref{sec:strategies}, we would start with parametrizing the box and then conclude that we would have to introduce one extra propagator as an ISP. Then the second loop would behave similarly to the pentagon integral discussed above, where we concluded that its five propagators were not really independent in pure 4D since they were related through the Gram determinant constraint of eq.~\eqref{eq:pentagram}. Yet where in the discussion above we used this constraint to write the pentagon in terms of sub-sectors, we may here use it to write the new ISP in terms of the four physical propagators on the pentagon side. And thus we conclude that no ISPs are needed for the finite pentabox after all, if one is interested only in the finite contribution.

Writing such a 4D Baikov representation for the finite pentabox explicitly will require solving eq.~\eqref{eq:pentagram} for the dot-product present only in the potential ISP (this could for instance be $k_2 \cdot p_4$). The most straight forward way to do this would use a tool such as the \texttt{Solve}-function of \texttt{Mathematica}, and for that reason it is not directly suitable for implementation in \texttt{BaikovPackage}. We will also not pursue it further here due to the size of the intermediate expressions.

Returning to the elliptic doublebox, it might seem that following the procedure described in sec.~\ref{sec:strategies} would introduce $N_{\text{extra}}=3$ extra propagators. Using, however, the fact discussed in app.~\ref{app:spacetime} that the last loop can never have $E$ higher that the spacetime dimensionality, we can show that the correct value is $N_{\text{extra}}=2$ for four spacetime dimensions. Since the elliptic doublebox is finite we can, additionally, consider the integral in pure $d=4$, and then the Gram determinant constraint of the type discussed above brings the number down by one more. This means that a 4D Baikov parametrization of the elliptic doublebox can be made with $N_{\text{extra}}=1$ as it was done in ref.~\cite{Frellesvig:2021vdl}; no more than are needed for the ordinary four-point doublebox.

Deriving this 4D Baikov parametrization in full generality, in a manner that does not require a generic \texttt{Solve}-step and thus allows for an implementation in a future version of \texttt{BaikovPackage}, would be a worthwhile future research project.

\subsection{Further discussion}

Putting the Baikov representation of eq.~\eqref{eq:standard} and the Lee--Pomeransky representation next to each other reveals something curious:
\begin{align}
I_{\{a\}} &= K \! \int_{\mathcal{C}} \mathcal{B}^{(d{-}E{-}L{-}1)/2} \, \Big( \prod_i x_i^{-a_i} \Big) \, \id^{n} x  \nonumber \\
&= \tilde{K} \! \int_0^{\infty} \!\! \mathcal{G}^{-d/2} \, \Big( \prod_i z_i^{a_i-1} \Big) \, \id^{P} z
\end{align}
Here $\mathcal{G} = \mathcal{U}+\mathcal{F}$ is the Lee--Pomeransky polynomial~\cite{Lee:2013hzt}, and we have absorbed the various constant prefactors into $K$ and $\tilde{K}$ to simplify the expressions.
We see that whereas the power of $\mathcal{B}$ goes as $d/2$ the power of $\mathcal{G}$ goes as $-d/2$, and while the powers of the Baikov variables $x$ go as $-a_i$ the powers of the Feynman parameters $z$ go as $a_i$. This seems to suggest some sort of duality between the two representations. It could be interesting to explore this in higher detail, perhaps by deriving one starting from the other\footnote{
%Some preliminary work in this direction is done in ref.~\cite{Chen:2023eqx}.
Since the original publication of this paper version 2 of ref.~\cite{Chen:2023eqx} has appeared, essentially doing this derivation.
}, as opposed to deriving each starting from the momentum representation as it is usually done.

We should also mention a different type of apparent duality between the two representations. This stems from the recently discovered fact~\cite{Lu:2024dsb} that in the language of \textit{relative twisted cohomology}~\cite{Caron-Huot:2021xqj, Brunello:2023rpq}, Feynman integrals in the Lee--Pomeransky representation are naturally expressed with \textit{delta-forms} in the integral basis, whereas in Baikov representation it is the dual forms that naturally contain the delta-forms.

There are some potential uses of parametric representations of Feynman integrals where the loop-by-loop Baikov representation fails. This includes the counting of master integrals (corresponding to the size of the twisted cohomology group describing the integral family) using the Lee--Pomeransky criterion~\cite{Lee:2013hzt}, which tends to over count when using the loop-by-loop, as opposed to the standard, Baikov representation. This problem carries over to the use of intersection theory starting from the loop-by-loop representation, since that approach requires a high degree of understanding of, and control over, the cohomological structure. The reason for this miscounting, seems to be that the counting is done as if the powers to which the $\mathcal{B}_i$ and $\mathcal{E}_i$ polynomials of eq.~\eqref{eq:loop-by-loop} are raised were generic free variables as opposed to the specific functions of $d$ given by that equation. Promoting these powers in that way defines what in ref.~\cite{Chen:2022lzr} is called the ``generalized loop-by-loop representation'', and it defines a larger class of objects, with more master integrals, than the Feynman integral family of the original problem. Whether or not there is a way to modify the loop-by-loop representation in such a way that the cohomological behaviour becomes correct, and which allows for the application of the full machinery of intersection theory, is an interesting open problem. 

The Baikov representation has shown itself to be an extremely useful tool for revealing mathematical structures underlying Feynman integrals. As discussed in the introduction, examples are the Baikov representation making the twisted cohomology structure of Feynman integrals explicit, as well as its ability to, through the maximal cut, reveal the geometric structures underlying the integrals. Yet, to the best of my knowledge, these tools have yet to be used by mathematicians, as opposed to mathematically minded physicists, interested in these mathematical structures. It is my belief and hope that some of the discussions in this paper, along with potential future progress on the issues discussed above such as a graph-based approach to Baikov parametrization or the purely 4D Baikov representation, will help to bridge that gap and make the Baikov representation as attractive to the mathematical community as it is in physics.

I hope with this publication to have clarified some aspects of the loop-by-loop Baikov representation that had not previously been explained in the literature, and with the added \texttt{BaikovPackage} (which can be downloaded at \url{https://github.com/HjalteFrellesvig/BaikovPackage}) to help make the Baikov representation, in its standard and loop-by-loop variants, more accessible.

\subsubsection*{Acknowledgments}

I would like to thank Costas Papadopoulos for developing the loop-by-loop Baikov representation with me back in 2016~\cite{Frellesvig:2017aai}. Furthermore I would like to thank the users of \texttt{BaikovPackage} through the ages, in particular Luca Mattiazzi, Federico Gasparotto, and Roger Morales, for much discussion on the package in its various stages, and on the loop-by-loop Baikov representation in general.
I would also like to thank Giacomo Brunello, Giulio Crisanti, Federico Gasparotto, Roger Morales, Costas Papadopoulos, Matthias Wilhelm, and Li Lin Yang, for reading through the manuscript in its draft stage and providing useful feedback.

This work was supported by the research grant 00025445 from Villum Fonden, and has received funding from the European Union’s Horizon 2020 research and innovation program under the Marie Sk{\l}odowska-Curie grant agreement No. 847523 ‘INTERACTIONS’.

\appendix

\section{Derivation of the Baikov representation}
\label{app:derivation}

The derivation of the Baikov representation from the momentum representation can by now be found in many places, e.g. refs.~\cite{Grozin:2011mt, Frellesvig:2017aai, Mastrolia:2018uzb, Frellesvig:2021vdl, Weinzierl:2022eaz}. Yet at various points in this paper we reference steps in that derivation, so we will nonetheless repeat it here in some detail.

\subsubsection*{The Baikov representation for one loop}

Our starting point we will be the momentum representation given by eq.~\eqref{eq:momentumrep}. In the case of one loop that is
\begin{align}
I_{\text{one-loop}} &= \int \frac{\id^d k}{i \pi^{d/2}} \frac{1}{\rho_1(k)^{a_1} \cdots \rho_{E{+}1}(k)^{a_{E{+}1}}}
\end{align}
Our first step will be to separate the integration into components that are respectively parallel and perpendicular to the subspace spanned by the $E$ independent momenta external to the integral, i.e. $\{p_1, \ldots, p_E\}$. That is $k = k_{||} + k_{\perp}$ where $k_{||} {\cdot} k_{\perp} = 0$. For the differential form this becomes\footnote{We might consider what happens in the special cases $E=0$ and $E=d$. $E=0$ corresponds to integrals of tadpole type, i.e. with no external momenta. For that case the following derivation goes through, as long as we define the determinant of a $0 \times 0$ matrix to be one, so $\mathcal{E}=1$. For $E=d$ the derivation fails, but that case is also explicitly unregulated by dimensional regularization. This is discussed further in sec.~\ref{sec:4D}. There can also be problems when $E=1$ and that is the topic of app.~\ref{app:masslesstwopoint}.}
\begin{align}
\text{d}^d k = \text{d}^E k_{||} \text{d}^{d-E} k_{\perp}
\label{eq:intsep}
\end{align}
Nothing in the integrand depends on $k_{\perp}$ other than through its square, so therefore it makes sense to split that part of the integral into a radial and an angular part\footnote{This variable change is akin to the cylindrical coordinates known from 3D Euclidean space. In cylindrical coordinates we have an axial coordinate $z$, a radial coordinate $r$, and an angular coordinate $\theta$, while here we have $E$ axial coordinates, one radial coordinate, and $d-E-1$ angular coordinates.} ($\id^n x = r^{n-1} \id r \id \Omega_{n-1}$) giving
\begin{align}
I_{\text{one-loop}} &= \frac{1}{i \pi^{d/2}} \int \frac{|k_{\perp}|^{d-E-1} \, \id^E k_{||} \, \id |k_{\perp}| \, \id \Omega_{d-E-1}}{\rho_1(k)^{a_1} \cdots \rho_{E{+}1}(k)^{a_{E{+}1}}}
\end{align}
We may then perform the angular integral using the hyper-spherical integral\footnote{It is the fact that eq.~\eqref{eq:hypersphere} yields sensible results for non-integer values of $n$ that allows us to make sense of dimensionally regulated Feynman integrals in the first place.}
\begin{align}
\int \! \id \Omega_{n{-}1} &= \frac{2 \pi^{n/2}}{\Gamma(n/2)}
\label{eq:hypersphere}
\end{align}
and combining with a change of variables to the square of the radial component $(k_{\perp}^2)$ gives over all
\begin{align}
I_{\text{one-loop}} &= \frac{1}{\Gamma((d-E)/2) i \pi^{E/2}} \int \frac{(k_{\perp}^2)^{(d-E-2)/2} \, \id^E k_{||} \, \id (k_{\perp}^2)}{\rho_1(k)^{a_1} \cdots \rho_{E{+}1}(k)^{a_{E+1}}}
\label{eq:intstep1}
\end{align}

Let us now introduce the variables $\varsigma_i = k \cdot p_i$ corresponding to $\varsigma_i = k_{||} \cdot p_i$. Doing the derivative with respect to a component of $k_{||}$ gives $\id \varsigma_i / \id k_{||}^j = {p_i}^j$. This corresponds to $d^E \varsigma = \det({p_i}^j) d^E k_{||}$. Since the matrix ${p_i}^j$ multiplied with its transpose gives the Gram matrix $G(p_1,\ldots,p_E)$, we now get
\begin{align}
\id^E k_{||} = \mathcal{E}^{-1/2} \id \varsigma_1 \cdots \id \varsigma_E
\label{eq:fancy1}
\end{align}
where
\begin{align}
\mathcal{E} = \det \! \big( G( p_1 , \ldots, p_E ) \big)
\end{align}
is the external Baikov polynomial. Correspondingly for the square of the perpendicular component we introduce the variable $\varsigma_0 = k^2$ corresponding to $\varsigma_0 = k_{||}^2 + k_{\perp}^2$ and thus $d \varsigma_0 / d (k_{\perp}^2) = 1$. This allows us to identify
\begin{align}
\id^E k_{||} \wedge \id (k_{\perp}^2) = \id^E k_{||} \wedge \id \varsigma_0
\label{eq:fancy2}
\end{align}
since remaining $k_{||}$-dependence in $\id \varsigma_0$ will annihilate when wedged with the $\id^E k_{||}$.

Let us now introduce the Baikov polynomial
\begin{align}
\mathcal{B} = \det \! \big( G( k, p_1 , \ldots, p_E ) \big)
\end{align}
We may write the loop momentum as $k = k_{||} + k_{\perp} = k_{\perp} + \sum_i \kappa_i p_i$ where the $\kappa_i$ are the components of $k_{||}$ along the individual $p_i$. Using then that a Gram determinant is invariant under linear shifts of the momenta within its vector space (corresponding to determinants being invariant under shifting rows/columns with multiples of other rows/columns), we may shift $k$ with $-k_{||}$ and get that $\mathcal{B} = \det( G( k_{\perp}, p_1 , \ldots, p_E ))$. Evaluating that determinant, we get that $\mathcal{B} = (k_{\perp}^2) \det( G( p_1 , \ldots, p_E ))$ corresponding to\footnote{Since a Gram determinant can be interpreted as the square of the volume of the parallelepiped spanned by its defining vectors, relations such as eq.~\eqref{eq:fancy3} can be interpreted geometrically. That is how this derivation is approached e.g. in ref.~\cite{Grozin:2011mt}.}
\begin{align}
(k_{\perp}^2) = \frac{\mathcal{B}}{\mathcal{E}}
\label{eq:fancy3}
\end{align}
We now have all the ingredients ready. Inserting eqs.~\eqref{eq:fancy2}, \eqref{eq:fancy1}, and \eqref{eq:fancy3} into the integral gives
\begin{align}
I_{\text{one-loop}} &= \frac{-i \, \pi^{-E/2} \, \mathcal{E}^{(E-d+1)/2}}{\Gamma((d-E)/2)} \int_{\mathcal{C}} \frac{\mathcal{B}(\varsigma)^{(d-E-2)/2} \, \id^{E+1} \varsigma}{\rho_1(\varsigma)^{a_1} \cdots \rho_{E{+}1}(\varsigma)^{a_{E+1}}}
\label{eq:oneloopinter}
\end{align}
It is the requirement that $k_{\perp}^2>0$ which translates into the integration region given by eq.~\eqref{eq:contourstandard}.
We may now do the final variable change from the $\varsigma_j$ to the Baikov variables $x_i = \rho_i$, which would result in the one-loop Baikov representation given by eq.~\eqref{eq:oneloop}. Let us, however, instead iterate the above to the multi-loop case.

\subsubsection*{The loop-by-loop Baikov representation}

For a multi-loop Feynman integral we have the momentum representation
\begin{align}
I &= \int \frac{\id^d k_1}{i \pi^{d/2}} \cdots \frac{\id^d k_L}{i \pi^{d/2}} \frac{N(k)}{\rho_1(k)^{a_1} \cdots \rho_P(k)^{a_P}}
\label{eq:momrepapp}
\end{align}
as given by eq.~\eqref{eq:momentumrep}. We realize that we may apply the parametrization of eq.~\eqref{eq:oneloopinter} to each of the $L$ individual loops. The only changes that would have to be made are replacing $E$ with the $E_i$ of the individual loops, and realizing that the $\mathcal{E}$ which in eq.~\eqref{eq:oneloopinter} was a constant that could be brought outside the integration, now would depend on the loop momenta of the remaining loops. We now also no longer necessarily have a correspondence between the number of propagators and the number of integration variables, corresponding to the introduction of irreducible scalar products at higher loop orders as discussed in sec.~\ref{sec:strategies}. Putting this together gives
\begin{align}
I &= \frac{(-i)^L \, \pi^{(L-n)/2} }{\prod_{l=1}^L \Gamma((d-E_l)/2)} \int_{\mathcal{C}} \frac{N(\varsigma) \, \prod_{l=1}^{L} \mathcal{E}_l(\varsigma)^{(E_l-d+1)/2} \, \mathcal{B}_l(\varsigma)^{(d-E_l-2)/2} \!}{\rho_1(\varsigma)^{a_1} \cdots \rho_P(\varsigma)^{a_P}} \, \id^{n} \varsigma
\label{eq:dotprodrep}
\end{align}
where we have kept the generic numerator\footnote{We have here assumed that that numerator can be written as $N(k) = N(\varsigma)$, meaning that it has no dependence on scalar products involving the loop momenta, that are not included among the $\varsigma$. For what to do otherwise, see app.~\ref{app:remaining}.} $N(\varsigma)$ from eq.~\eqref{eq:momrepapp} and inserted
\begin{align}
n = \sum_{l=1}^L (E_l + 1) = L + \sum_{l=1}^L E_l
\end{align}
as also given by eq.~\eqref{eq:nlbl}.
Now all that is left to derive the loop-by-loop Baikov representation is to perform the final variable change to the Baikov variables, giving
\begin{align}
I &= \frac{\mathcal{J} \, (-i)^L \, \pi^{(L-n)/2} }{\prod_{l=1}^L \Gamma((d-E_l)/2)} \int_{\mathcal{C}} \frac{N(x) \, \prod_{l=1}^{L} \mathcal{E}_l(x)^{(E_l-d+1)/2} \, \mathcal{B}_l(x)^{(d-E_l-2)/2} \!}{x_1^{a_1} \cdots x_P^{a_P}} \, \id^{n} x
\label{eq:loop-by-loop-app}
\end{align}
where $\mathcal{J}$ is the Jacobian for that variable change. That Jacobian will be given as $\mathcal{J} = \text{det}(\id \varsigma_i / \id x_j)$, but it is more convenient to compute it as the inverse Jacobian of the reverse change, i.e.
\begin{align}
\mathcal{J} &= 1 / \text{det} ( Y ) \qquad \text{with} \qquad Y_{ij} = \frac{\id x_i}{\id \varsigma_j}
\end{align}
For most types of propagators, a $\varsigma_j$ of the type $k^2$ will give a column in $Y$ containing $0$ or $\pm 1$, while a $\varsigma_j$ of the form $k \cdot q$ will give $0$ or $\pm 2$. As there will be $L$ of the former and thus $n-L$ of the latter, we expect, using properties of determinants, that $\mathcal{J} = \pm 2^{L-n}$.

Replacing the numerator $N(x)$ of eq.~\eqref{eq:loop-by-loop-app} with one of the monomials therein, we reproduce eq.~\eqref{eq:loop-by-loop}.

\subsubsection*{The standard Baikov representation}

To derive the standard Baikov representation we will start from eq.~\eqref{eq:dotprodrep}, and then take each loop to depend on all the momenta external to the whole integral, along with all the remaining loop momenta as also discussed at the end of sec.~\ref{sec:loop-by-loop}. In that case $E_l = E + L - l$ and we get
\begin{align}
n \; = \; L + \sum_{l=1}^L (E + L - l) \; = \; L (L+1)/2 + EL
\end{align}
as given by eq.~\eqref{eq:nstd}.
Furthermore we get that $\mathcal{E}_l$ = $\mathcal{B}_{l+1}$ and that the powers to which those polynomials are raised become each others negatives. Thus these polynomials will cancel \textit{telescopically} two and two, leaving only $\mathcal{B}_1$ and $\mathcal{E}_L$ which we will rename to $\mathcal{B}$ and $\mathcal{E}$. Changing variables from $\varsigma$ to $x$, we get
\begin{align}
I &= \frac{\mathcal{J} \, (-i)^L \, \pi^{(L-n)/2} \, \mathcal{E}^{(E-d+1)/2}}{\prod_{l=1}^L \Gamma \big( (d{-}E{-}L{+}l)/2 \big)} \int_{\mathcal{C}} \frac{N(x) \, \mathcal{B}(x)^{(d-E-L-1)/2} \!}{x_1^{a_1} \cdots x_P^{a_P}} \, \id^{n} x
\end{align}
where we have used the fact that $\mathcal{E}$ can be taken outside the integral, since it is constructed only from momenta external to the last loop in the integration ordering and therefore has no loop momentum dependence and thus no $x$-dependence. Once again replacing the numerator $N(x)$ with one of the monomials therein, and changing the index on the gamma functions, we reproduce the standard Baikov representation, eq.~\eqref{eq:standard}. This completes our derivation of the two versions of the Baikov representation starting from the momentum representation.

\section{What \texttt{BaikovPackage} cannot help you with}
\label{app:cannot}

There are Feynman integrals that cannot be Baikov parametrized directly, and where, for that reason, the functions in \texttt{BaikovPackage} will fail or return an error. In this appendix we will discuss various classes of such integrals and how to make valid parametrizations for them.

\subsection{Massless two-point kinematics}
\label{app:masslesstwopoint}

\begin{figure}
\centering
\includegraphics[width=13cm]{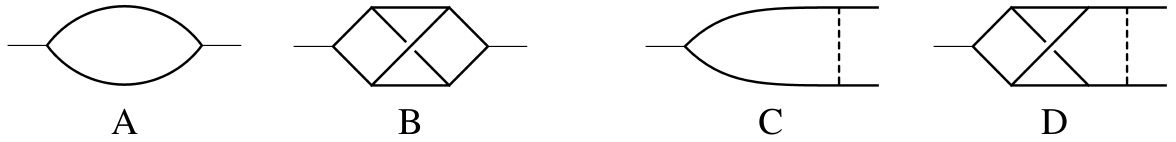}
%\vspace{0mm}
\caption{ }
\label{fig:partf1}
\end{figure}

The first class of integrals we will discuss are integrals with two-point kinematics, for which the external leg is massless, i.e. $p^2 = 0$. Two examples of such integrals are depicted in figs.~\ref{fig:partf1}A and B. The indication of the Baikov representation failing if directly applied to such diagrams, is the $\mathcal{E}$ in eq.~\eqref{eq:standard} which was defined as the Gram determinant of the full set of independent external momenta. With the kinematics given here that gives $\mathcal{E} = p^2 = 0$, making eq.~\eqref{eq:standard} invalid\footnote{One might consider what step in the derivation of the Baikov representation given in app.~\ref{app:derivation} becomes invalid in this special case. The mistake is introduced at the very beginning since the decomposition $k = k_{||} + k_{\perp}$ (where here $k_{||} = \alpha p$) becomes impossible for a generic $k$ if we require both that $k_{||} \cdot k_{\perp} = 0$ and $k_{||}^2 = 0$.}. For the loop-by-loop Baikov representation the same problem will appear when parametrizing the last loop.

The solution is to embed the integral in a larger integral sector with less trivial external kinematics. Examples of such larger sectors are shown in figs.~\ref{fig:partf1}C and D. For loop-by-loop this will always only introduce one extra Baikov variable into the problem, that which is shown as the dotted line in figs.~\ref{fig:partf1}C and D. In the one-loop case this is the only example in which an ISP is needed.

As a specific example let us look at the one-loop bubble $I^{\text{bub}}_{\nu_1 \nu_2}$ of fig.~\ref{fig:partf1}A with the two propagators
\begin{align}
P_1 \,=\, k^2 - m_1^2 \;, \qquad P_2 \,=\, (k-p)^2 - m_2^2
\end{align}
and with $p^2 = 0$. To Baikov parametrize this integral we have to embed it into a triangle sector $I^{\text{tri}}$ by introducing a third propagator
\begin{align}
P_3 \,=\, (k+q)^2 
\end{align}
that will play the role of an ISP. This triangle sector is shown in fig.~\ref{fig:partf1}C. The kinematics has to be chosen as sensible three-point kinematics, for example $q^2=0$ and $(p + q)^2 = s$. We then get the embedding
\begin{align}
I^{\text{bub}}_{\nu_1 \nu_2} \, = \, I^{\text{tri}}_{\nu_1 \nu_2 0} \, = \, \frac{\mathcal{J}_{\text{tri}} \, \mathcal{E}_{\text{tri}}^{(3-d)/2}}{i \pi \Gamma((d-2)/2)} \int_{\mathcal{C}} \frac{\mathcal{B}_{\text{tri}}^{(d-4)/2}}{x_1^{\nu_1} \, x_2^{\nu_2}} \, \id^3 x
\end{align}
The apparent dependence (through $\mathcal{E}_{\text{tri}}$ and $\mathcal{B}_{\text{tri}}$) on the spurious kinematic variable $s$ should drop out when evaluating the bubble integral.

\subsection{Linearly dependent propagators}
\label{app:linearlydependent}

\begin{figure}
\centering
\includegraphics[width=14cm]{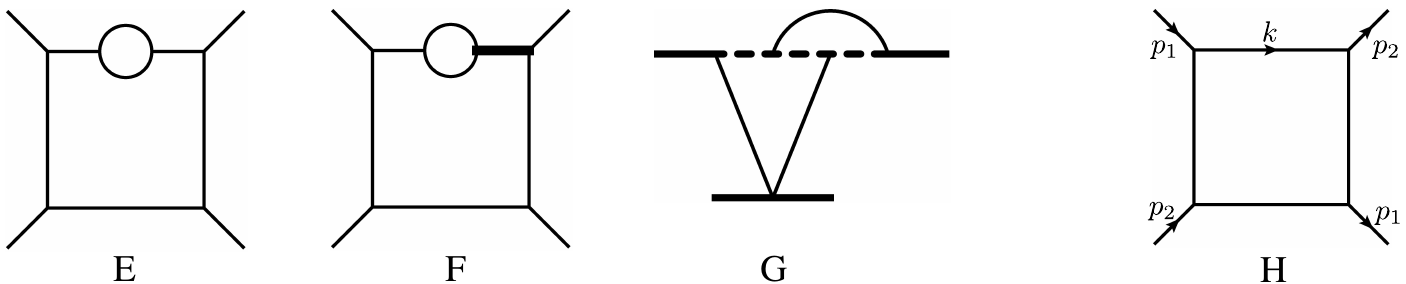}
%\vspace{0mm}
\caption{ }
\label{fig:partf2}
\end{figure}

Another type of integrals that cannot be directly Baikov parametrized are those for which the propagators are not linearly independent. This can happen when the external kinematics is somehow degenerate, as we will discuss in the following sections, but it can also happen generically. Examples of this are shown in fig.~\ref{fig:partf2}E--G. For the ``flying squirrel'' integrals in fig.~\ref{fig:partf2}E and F the loop momenta on each side of the bubble subgraph are equal. Fig.~\ref{fig:partf2}E is meant to illustrate the case where the two propagators on each side are completely identical. If $k_1$ is the momentum entering the bubble subgraph, we may parametrize the family as
\begin{align}
P_1 &= k_2^2 - \mu_1 \,, & P_2 &= (k_1{-}k_2)^2 - \mu_2 \,, & P_3 &= (k_1{+}p_4)^2 - \mu_3 \,, \nonumber \\
P_4 &= (k_1{-}p_1{-}p_2)^2 - \mu_4 \,, & P_5 &= (k_1{-}p_1)^2 - \mu_5 \,, & P_6 &= k_1^2 - \mu_6 \,, \label{eq:turtle}
\end{align}
where the $\mu_i$ denote generic squared particle masses. In that case the correct prescription is to treat that sixth propagator as doubled, so the integral depicted in fig.~\ref{fig:partf2}E would be written as $I_{111112}$.

This brings us to the integral in fig.~\ref{fig:partf2}F. That also has the incoming and outgoing momentum on the bubble subgraph being equal, but the masses carried by the two propagators are different:
\begin{align}
P_{6;1} &= k_1^2 - m_1^2 & P_{6;2} &= k_1^2 - m_2^2
\end{align}
The set of independent dot-products has six members, $\{k_1^2, k_2^2, k_1{\cdot}k_2, k_1{\cdot}p_1, k_1{\cdot}p_2, k_1{\cdot}p_4\}$, making inverting the seven propagator system impossible. We may define two integral families fam$1$ and fam$2$ as in eq.~\eqref{eq:turtle}, except that in fam$1$ $\mu_6 = m_1^2$ while in fam$2$ $\mu_6 = m_2^2$, where each of these families are valid on their own. We are now faced with two options:

The first option we will consider is to choose one of the families, say fam1, and do the Baikov parametrization there. This will treat the remaining propagator, $P_{6;2}$, differently from the rest, in that it will not be promoted to an integration variable but rather treated as a generic multiplicative factor similarly to how one would treat a numerator factor $N(k)$. The result is
\begin{align}
I_{\text{fig.\ref{fig:partf2}F}} &= K \! \int_{\mathcal{C}} \frac{\mathcal{B}^{(d{-}6)/2}_{\text{fam}1} \, \id^{n} x}{x_1 x_2 x_3 x_4 x_5 x_6 (x_6 + m_1^2 - m_2^2)}
\label{eq:falsebaikov}
\end{align}
where we have absorbed the constant prefactors into the generic $K$. A similar expression could be constructed for the loop-by-loop parametrization. Cutting the six propagators of fam1 gives $P_{6;2} = m_1^2 - m_2^2$, so with this expression we immediately realize that the maximal cut of integrals such as that of fig.~\ref{fig:partf2}F will always be $0$; it is impossible to find a contour that simultaneously cuts $P_{6;1}$ and $P_{6;2}$. Since not all propagators have been promoted to integration variables we are, however, not going to consider eq.~\eqref{eq:falsebaikov} a genuine Baikov parametrization.

The other option is to partial fraction the product of $P_{6;1}$ and $P_{6;2}$ as
\begin{align}
\frac{1}{(k_1^2 - m_1^2) (k_1^2 - m_2^2)} = \frac{1}{m_1^2 - m_2^2} \left( \frac{1}{k_1^2 - m_1^2} - \frac{1}{k_1^2 - m_2^2} \right)
\end{align}
and then we get that the integral in fig.~\ref{fig:partf2}F can be expressed as
\begin{align}
I_{\text{fig.\ref{fig:partf2}F}} &= \frac{1}{m_1^2 - m_2^2} \left( I^{\text{fam}1}_{111111} - I^{\text{fam}2}_{111111} \right)
\label{eq:squirrelpartial}
\end{align}
where the two families will have to be Baikov parametrized individually. Also from this expression we realize that integrals of this flying squirrel type will always be reducible to subsectors.

These two options will always arise when a Feynman integral has linearly dependent propagators. Unlike the example in sec.~\ref{app:masslesstwopoint} it is not just a direct Baikov parametrization that will fail, it is also impossible to directly perform e.g. an IBP decomposition of a Feynman integral with linearly dependent propagators; a rewriting akin to eq.~\eqref{eq:squirrelpartial} will have to be done also for that purpose. The same will hold true for the remaining examples in this appendix.

There are types of Feynman integrals where some propagators are \textit{linear}. This happens in the soft limit of Feynman integrals as derived in the expansion by regions~\cite{Beneke:1997zp} and specifically in the post-Minkowskian approach~\cite{Bern:2019nnu} to classical gravity. For such integrals there are less straightforward ways in which propagators can be linearly dependent. An example is depicted in fig.~\ref{fig:partf2}G where the thick dotted line depicts the linearized propagators.
Specifically that integral may be parametrized with the six propagators
\begin{align}
P_1 &= k_1^2 \,, & P_2 &= (k_1{+}q)^2 \,, & P_3 &= (k_1{-}k_2)^2 \,, \nonumber \\
P_4 &= 2 k_1 {\cdot} u \,, & P_{5;1} &= 2 k_2 {\cdot} u \,, & P_{5;2} &= 2 (k_2{-}k_1) {\cdot} u \,, \label{eq:gravity}
\end{align}
where the kinematics is such that $u^2 = 1$, $q^2 = t$, $u \cdot q = 0$, a fact that will not be used in the following. The three propagators $P_4$, $P_{5;1}$, and $P_{5;2}$ together only contain two scalar products and as such they are not linearly independent. Once again we can do partial fractioning
\begin{align}
\frac{1}{P_4 P_{5;1} P_{5;2}} = \frac{1}{P_4^2} \left( \frac{1}{P_{5;2}} - \frac{1}{P_{5;1}} \right)
\end{align}
so we see that if we define two integral families in the obvious way, we may write
\begin{align}
I_{\text{fig.\ref{fig:partf2}G}} &= I^{\text{fam}2}_{11121} - I^{\text{fam}1}_{11121}
\end{align}
where again the two families will have to be Baikov parametrized separately. Also in this case it would be possible to choose one parametrization and write the remaining propagator in terms of that, as it was done in eq.~\eqref{eq:falsebaikov}, but we will not pursue this further.
It is the existence of this kind of non-trivial relations in the presence of linear propagators, which enables the relations between the maximal cuts of for instance certain planar and non-planar integrals in the post-Minkowskian expansion, as discussed e.g. in ref.~\cite{Frellesvig:2024zph}.

\subsection{Degenerate kinematics}
\label{app:degenerate}

Another class of Feynman integrals to which the Baikov representation is not directly applicable are those for which the kinematics is \textit{degenerate}, by which we will mean that the external momenta obey relations in addition to that of over-all momentum conservation. This can happen in soft or collinear limits, and also for the cut Feynman integrals that arise when expressing phase-space integrals using reverse unitarity~\cite{Anastasiou:2002yz}. Since such degenerate kinematics will cause the propagators to become linearly dependent, this is a special case of the problem discussed in app.~\ref{app:linearlydependent}. It is, however, worth discussing separately.

As an example\footnote{This is the same example as that discussed, for the same reason, in sec.~2.5.5 of ref.~\cite{Weinzierl:2022eaz}.} let us consider the one-loop box in the collinear limit $p_4 \rightarrow - p_2$, which is shown in fig~\ref{fig:partf2}H. In that case we get 
\begin{align}
P_1 &= k^2 - \mu_1 \,, & P_2 &= (k{-}p_1)^2 - \mu_2 \,, & P_3 &= (k{-}p_1{-}p_2)^2 - \mu_3 \,, & P_4 &= (k_1{-}p_2)^2 - \mu_4 \,,
\end{align}
where we have allowed for generic squared particle masses $\mu_i$. Let us call this integral family $I_{\text{deg-box}}$. Multiplying out, we see that only three dot-products $\{ k^2, k \cdot p_1, k \cdot p_2 \}$ involving the loop momentum are present, so clearly this mapping is not invertible. As in app.~\ref{app:linearlydependent} we now have two options for how to proceed:

One option is to treat one of the propagators, e.g. $P_4$, differently from the rest by treating it as an ordinary multiplicative factor. In that case the parametrization proceeds exactly as it would for a triangle containing only $P_1$ to $P_3$. The result is
\begin{align}
I_{\text{deg-box}} &= \frac{\mathcal{J}_{\text{tri}} \, \mathcal{E}_{\text{tri}}^{(3-d)/2}}{i \pi \Gamma((d-2)/2)} \int_{\mathcal{C}} \frac{\mathcal{B}_{\text{tri}}^{(d-4)/2} \, \id^3 x}{x_1^{\nu_1} x_2^{\nu_2} x_3^{\nu_3} P_4(x)^{\nu_4}}
\label{eq:degbox}
\end{align}
The expression for $P_4(x)$ (as well as $\mathcal{E}_{\text{tri}}$ and $\mathcal{B}_{\text{tri}}$) will depend on the exact kinematics, but in the simplest case where 
\begin{align}
\mu_1 = \mu_2 = \mu_3 = \mu_4 = 0 \,, \qquad p_1^2 = p_2^2 = 0 \,, \qquad (p_1+p_2)^2 = s \,,
\label{eq:degkin}
\end{align}
we get
\begin{align}
P_4(x) &= x_1 - x_2 + x_3 - s
\end{align}
Cutting the propagators $P_1$--$P_3$ will make $P_4$ a constant, so we see that cutting all four simultaneously is an impossibility. From that fact we instantly see that this degenerate box will not be a master integral since it has a vanishing maximal (i.e. four-propagator) cut. 

The second option is to partial fraction the product of the four propagators. This is most easily done by writing the $1$ in the numerator of $1/(P_1 P_2 P_3 P_4)$ in terms of the four propagators
\begin{align}
1 &= b_1 P_1 + b_2 P_2 + b_3 P_3 + b_4 P_4
\label{eq:oneone}
\end{align}
and then fitting $b_1$ to $b_4$ to make the coefficients of the three dot-products $\{ k^2, k \cdot p_1, k \cdot p_2 \}$ equal zero, and the constant remainder equal to one. For the kinematics given by eq.~\eqref{eq:degkin} we get
\begin{align}
b_1 = \frac{1}{s} \,,\qquad b_2 = -\frac{1}{s} \,,\qquad b_3 = \frac{1}{s} \,,\qquad b_4 = -\frac{1}{s} \,,
\end{align}
and inserting that gives the specific decomposition of the integral
\begin{align}
I^{\text{deg-box}}_{1111} &= \frac{1}{s} \left( I^{\text{deg-box}}_{0111} - I^{\text{deg-box}}_{1011} + I^{\text{deg-box}}_{1101} - I^{\text{deg-box}}_{1110} \right)
\end{align}
where we now may Baikov parametrize each of the terms on the right-hand side individually as if they were triangles.

It may happen that a kinematic constraint is given, not in terms of the external momenta themselves, but rather in terms of the Mandelstam variables and masses, e.g. of the form $t \rightarrow -s$. To check if such a limit causes the set of propagators to become linearly dependent, one might compute the Gram determinant $\mathcal{E}$ as given by eq.~\eqref{eq:EandB}. If that determinant remains finite, this limit will cause no problems for the Baikov parametrization. If, however, $\mathcal{E}$ vanishes in the limit, we know that the limit has reduced the dimensionality of the space spanned by the external momenta, causing $E$ to decrease. In that case the limit will have to be translated into a limit in terms of the external momenta, which then can be treated as discussed above. How to perform that translation, is identical to the way it is done for the specific type of kinematics discussed in the following section.

\subsection{Kinematic constrains from finite spacetime dimensionality}
\label{app:spacetime}

\begin{figure}
\centering
\includegraphics[width=5cm]{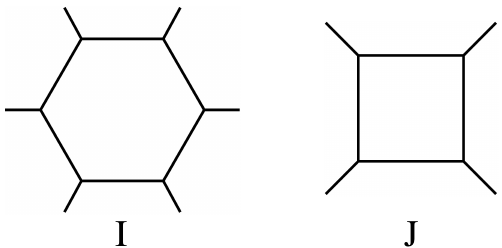}
%\vspace{0mm}
\caption{ }
\label{fig:partf3}
\end{figure}

Another class of Feynman integrals that cannot be Baikov parametrized directly are those where the finite dimensionality of spacetime makes it impossible for the external momenta to be generic. This is a special case of the degenerate kinematics discussed in app.~\ref{app:degenerate}, but it is still worth discussing it separately. That dimensional restriction imposes a Gram determinant constraint which says that
\begin{align}
\det \! \big( G ( p_1, \cdots , p_n ) \big) &= 0 \qquad \text{whenever} \qquad n>d_{\text{spacetime}}
\label{eq:gramconstraint}
\end{align}
where $d_{\text{spacetime}}$ is taken to be an integer. In $d_{\text{spacetime}}=4$ this constraint has an effect starting at six-point, as illustrated by the hexagon integral shown of fig.~\ref{fig:partf3}I, where it tells you that you cannot have five independent external momenta, so taking $E=5$ for such cases would be incorrect. It is easy to see that the naive expression for a Baikov parametrization (with $E=5$) cannot hold in this case, since the polynomial $\mathcal{E}$ of eq.~\eqref{eq:standard} is exactly the Gram determinant of eq.~\eqref{eq:gramconstraint}.
The fact that the five momenta cannot be linearly independent in four dimensions can be made explicit by writing
\begin{align}
p_5 = \tilde{a}_1 p_1 + \tilde{a}_2 p_2 + \tilde{a}_3 p_3 + \tilde{a}_4 p_4 
\end{align}
which will make the Gram determinant constraint of eq.~\eqref{eq:gramconstraint} be satisfied automatically. This makes explicit the fact that there only can be five independent scalar products involving the loop momentum $\{ k^2, k \cdot p_1, k \cdot p_2, k \cdot p_3, k \cdot p_4 \}$, so the six propagators cannot be linearly independent. We could continue with this example, but since six-point kinematics is rather involved, let us redirect our attention to a different problem, that of the four-point box in two spacetime dimensions.

The four-point box integral is depicted in fig.~\ref{fig:partf3}J. In the internally massless case we may write the four propagators as
\begin{align}
P_1 &= k^2 \,, & P_2 &= (k{+}p_1)^2 \,, & P_3 &= (k{+}p_1{+}p_2)^2 \,, & P_4 &= (k{+}p_1{+}p_2{+}p_3)^2 \,.
\end{align}
From momentum conservation we have that $p_4 = -p_1-p_2-p_3$, but if we embed the integral in two spacetime dimensions, we get the additional constraint
\begin{align}
\det \! \big( G ( p_1, p_2, p_3 ) \big) &= 0
\end{align}
from eq.~\eqref{eq:gramconstraint}. Introducing the standard Mandelstam variables $s=(p_1{+}p_2)^2$, $t=(p_2{+}p_3)^2$, we now get that $s$ and $t$ no longer are allowed to vary independently. We will in the following consider $s$ to be the independent variable.
We may make the two dimensional restriction explicit by writing
\begin{align}
p_3 = a_1 p_1 + a_2 p_2
\end{align}
We can then fit $a_1$ and $a_2$ by imposing the physical values for $p_3^2$ and $p_4^2$. We will now as an example consider the specific kinematics called ``two mass hard'', in which
\begin{align}
p_1^2 = p_2^2 = m^2 \,,\quad p_3^2=p_4^2 = 0
\end{align}
This corresponds to two solutions for the $a_i$ which are given as
\begin{align}
a_1 = -\frac{(4 m^2 {-} s) \pm \sqrt{-s(4m^2 {-} s)}}{2 (4 m^2 {-} s)} \,,\quad a_2 = -\frac{(4 m^2 {-} s) \mp \sqrt{-s(4m^2 {-} s)}}{2 (4 m^2 {-} s)} \,,
\end{align}
corresponding to two distinct momentum configurations giving rise to the specified kinematics.

We may then perform the partial fractioning of the integrand $1/(P_1 P_2 P_3 P_4)$. This is easiest done as discussed in eq.~\eqref{eq:oneone}, i.e. by writing the $1$ in the numerator as
\begin{align}
1 = b_1 P_1 + b_2 P_2 + b_3 P_3 + b_4 P_4
\end{align}
and then fit the $b_i$s by the requirement that the coefficients of each of $k^2$, $k {\cdot} p_1$, and $k {\cdot} p_2$ should be zero, and the constant term should be one.
Doing so gives the partial fractioning
\begin{align}
\frac{1}{P_1 P_2 P_3 P_4} = \frac{b_1}{P_2 P_3 P_4} + \frac{b_2}{P_1 P_3 P_4} + \frac{b_3}{P_1 P_2 P_4} + \frac{b_4}{P_1 P_2 P_3}
\end{align}
with the $b$-values
\begin{align}
b_1 = \frac{s \mp r}{2 m^2 s} \,,\quad b_2 = \frac{2m^2 - s \pm r}{2 m^4} \,,\quad b_3 = \frac{s \mp r}{2 m^2 s} \,,\quad b_4 = \frac{r^2 \pm r (2 m^2 - s)}{2 m^4 s} \,,
\end{align}
with $r = \sqrt{{-}s(4m^2{-}s)}$. We may now Baikov parametrize each of the four triangles individually, as in the previous examples.

\section{The remaining dot-products as numerator factors}
\label{app:remaining}

In this appendix we will discuss Feynman integrals in which the numerator contains scalar products that would not be included in a loop-by-loop Baikov parametrization of that integral.

The approach discussed in the following was first introduced in (sec.~11.1 of) ref.~\cite{Frellesvig:2019kgj} but will here be described in higher detail. It is inspired by the Passarino--Veltman reduction~\cite{Passarino:1978jh} used for integrand reduction, and we should also mention the connection to similar ideas recently used at the level of IBP reduction in ref.~\cite{Chestnov:2024mnw}.

Specifically the aim is to derive a replacement rule that can be applied to such extra scalar products, and which therefore can be used to replace any scalar numerator $N(k)$ of the original integral in momentum space with an $N(x)$, as it is given in eq.~\eqref{eq:loop-by-loop-app}, containing only the variables of the loop-by-loop Baikov representation.

As an illustrative one-loop example we will consider the triangle-integral
\begin{align}
\int \! \frac{\id^d k}{i \pi^{d/2}} \, \frac{N(k)}{P_1^{\nu_1} P_2^{\nu_2} P_3^{\nu_3}}
\end{align}
where
\begin{align}
P_1 = k^2 \,, \qquad P_2 = (k-p_1)^2 \,, \qquad P_3 = (k+p_2)^2 \,.
\label{eq:triprobs}
\end{align}
Expanding the propagators will introduce the scalar products
\begin{align}
\left\{ k^2 \,,\;\; k \cdot p_1 \,,\;\; k \cdot p_2 \right\}
\end{align}
which can be mapped back to the propagators in an invertible fashion, so it is clear what to do with any instances of these three scalar products if they appear in $N(k)$. But what do we do if a new scalar product\footnote{In this appendix we will make no assumptions about momentum conservation for the set of $p_i$, in order to not restrict the discussion to specific $n$-point kinematics.}, such as $k \cdot p_3$, appears in $N(k)$? One obvious solution would be to promote it to a new ``extra'' propagator $P_4$ which after the Baikov parametrization would become an extra integration variable that would not be present otherwise. Doing so would, however, remove some of the benefits of using the loop-by-loop approach. This supposed inability of the loop-by-loop Baikov representation to handle such cases is sometimes regarded as a weakness of that approach, but this is a misconception since it is possible to handle numerators of that type in a systematic fashion without introducing extra integration variables. We will now show how that can be done.

Let us first consider the case where $N(k) = (k \cdot p_3)$. The first step will be to write
\begin{align}
p_3^{\mu} &= a_1 p_1^{\mu} + a_2 p_2^{\mu} + b \eta^{\mu}
\label{eq:p3exp}
\end{align}
where the vector $\eta^{\mu}$ is defined to be perpendicular to $p_1^{\mu}$ and $p_2^{\mu}$, i.e. $p_1{\cdot}\eta = p_2{\cdot}\eta = 0$.
The coefficients $a_1$ and $a_2$ can be found by contracting eq.~\eqref{eq:p3exp} with ${p_1}_{\mu}$ and ${p_2}_{\mu}$ and solving the resulting equation system. The coefficient $b$ can likewise be found by contracting with ${p_3}_{\mu}$, but doing that is not needed for what follows.
The motivation for the rewriting of eq.~\eqref{eq:p3exp} is that $N(k) = (k \cdot p_3)$ now may be written in terms of $(k \cdot p_1)$ and $(k \cdot p_2)$ which can be expressed in terms of the Baikov variables of eq.~\eqref{eq:triprobs}, along with $(k \cdot \eta)$ which may easily be seen to integrate to zero since
\begin{align}
\int \frac{\id^d k}{i \pi^{d/2}} \frac{\eta_{\mu} k^{\mu}}{P_1^{\nu_1} P_2^{\nu_2} P_3^{\nu_3}} \, = \, \eta_{\mu} (J_1 p_1^{\mu} + J_2 p_2^{\mu}) \, = \, 0
\end{align}
where we have used the fact that a vector-valued Feynman integral can only evaluate to a linear combination of vectors occurring in the problem.
If we have the kinematics
\begin{align}
p_i^2 = 0 \,,\quad p_1 \cdot p_2 = s/2 \,,\quad p_2 \cdot p_3 = t/2 \,,\quad p_1 \cdot p_3 = u/2 \,,
\label{eq:trikinematics}
\end{align}
we get
\begin{align}
a_1 = \frac{t}{s} \, , \quad a_2 = \frac{u}{s} \,,\quad (k \cdot p_1) = \frac{x_1-x_2}{2} \,,\quad (k \cdot p_2) = \frac{x_3-x_1}{2} \,.
\end{align}
This means that we may perform the replacement
\begin{align}
N(k) = (k \cdot p_3) \quad \rightarrow \quad N(x) = a_1 (k \cdot p_1) + a_2 (k \cdot p_2) = \frac{(t-u) x_1 - t x_2 + u x_3}{2 s}
\end{align}
In other words loop-by-loop parametrizing the integral with $N(k) = (k \cdot p_3)$ can be done by multiplying the integrand of the ordinary loop-by-loop Baikov parametrization with the $N(x)$ given above.

For the next example we will consider $N(k) = (k \cdot p_3)^2$. Attempting a rewriting similar to the previous case, we may write $p_3^{\mu} p_3^{\nu}$ as the appropriate ``square'' of eq.~\eqref{eq:p3exp}:
\begin{align}
p_3^{\mu} p_3^{\nu} &= \left( a_1 p_1^{\mu} + a_2 p_2^{\mu} + b \eta^{\mu} \right) \left( a_1 p_1^{\nu} + a_2 p_2^{\nu} + b \eta^{\nu} \right)
\end{align}
While not incorrect, doing so would be misleading since the term $(k \cdot \eta)^2$ does not integrate to zero as the integral over $k^{\mu} k^{\nu}$ can yield a term proportional to $g^{\mu \nu}$, and the resulting $\eta^2$ will not be zero in general. The correct approach is to introduce the $g^{\mu \nu}$ in the basis explicitly, replacing the $\eta^{\mu} \eta^{\nu}$ term. In other words we will use the expansion
\begin{align}
p_3^{\mu} p_3^{\nu} &= a_1 p_1^{\mu} p_1^{\nu} + a_2 p_1^{\mu} p_2^{\nu} + a_3 p_2^{\mu} p_1^{\nu} + a_4 p_2^{\mu} p_2^{\nu} + a_5 g^{\mu \nu} + b_1 \eta^{\mu} p_1^{\nu} + b_2 \eta^{\mu} p_2^{\nu} + b_3 p_1^{\mu} \eta^{\nu} + b_4 p_2^{\mu} \eta^{\nu}
\label{eq:p3sq}
\end{align}
The coefficients $a_1$ to $a_5$ may now be extracted by contracting eq.~\eqref{eq:p3sq} with (cotensors to) the members of the basis
\begin{align}
\big\{ p_1^{\mu} p_1^{\nu} \,,\;\; p_1^{\mu} p_2^{\nu} \,,\;\; p_2^{\mu} p_1^{\nu} \,,\;\; p_2^{\mu} p_2^{\nu} \,,\;\; g^{\mu \nu} \big\}
\label{eq:basis}
\end{align}
and solving the resulting equation system. The four $b_i$-terms are irrelevant since they multiply tensors that will integrate to zero as in the example above.
The result is thus that we may perform the replacement
\begin{align}
N(k) = (k \cdot p_3)^2 \quad \rightarrow \quad N(x) = a_1 (k \cdot p_1)^2 + (a_2+a_3) (k \cdot p_1) (k \cdot p_2) + a_4 (k \cdot p_2)^2 + a_5 k^2
\end{align}
which with the kinematics given by eq.~\eqref{eq:trikinematics} corresponds to 
\begin{align}
N(x) &= \frac{(d{-}2) t^2 (x_1{-}x_2)^2 + (d{-}2) u^2 (x_1{-}x_3)^2 - 2 d t u (x_1{-}x_2)(x_1{-}x_3) - 4 s t u x_1}{4 (d{-}2) s^2}
\end{align}

Usually a basis similar to that given by eq.~\eqref{eq:basis} is the best way of thinking about this tensorial expansion, so we may in general write
\begin{align}
\text{tensor} \; &= \; \!\!\!\!\! \sum_{i: \, \beta_i \in \text{basis}} \!\!\!\!\! a_i \beta_i \;\; + \;\; \text{spurious}
\end{align}
where ``spurious'' refers to terms that will integrate to zero, which corresponds to terms containing an odd number of $\eta$s. When constructing the basis, starting from the appropriate power of the equivalent to eq.~\eqref{eq:p3exp} is a valid approach, keeping in mind that terms with even powers of $\eta$ have to have the $\eta$s replaced by $g^{\mu \nu}$. Starting at the fourth tensorial order\footnote{One might imagine also needing the Levi--Civita tensor $\varepsilon^{\mu_1 \mu_2 \mu_3 \mu_4}$ in the basis at this order. This could be done, but would be equivalent to allowing for objects of the form $\text{Tr}(\gamma_5 {k \!\!\! /} {p \!\!\! /}_1 {p \!\!\! /}_2 {p \!\!\! /}_3)$ in the numerator, as opposed to just scalar products. That goes beyond the scope of this publication and of \texttt{BaikovPackage}.}, more than one such $g^{\mu \nu}$-term is needed to replace four powers of $\eta$, in particular
\begin{align}
\eta^{\mu_1} \eta^{\mu_2} \eta^{\mu_3} \eta^{\mu_4} \; \rightarrow \; \big\{ g^{\mu_1 \mu_2} g^{\mu_3 \mu_4} ,\; g^{\mu_1 \mu_3} g^{\mu_2 \mu_4} ,\; g^{\mu_1 \mu_4} g^{\mu_2 \mu_3} \big\}
\end{align}
and likewise at each new even tensorial power new structures will appear along similar lines.

We can derive the size of the linear system that has to be solved. For a loop with $E$ independent external legs, a numerator of tensorial degree $\tau$ requires
\begin{align}
N_{\text{terms}} &= \sum_{i=0}^{\left\lfloor \tfrac{\tau}{2} \right\rfloor} \frac{\tau! \, E^{\tau-2i}}{(\tau-2i)! \, i! \, 2^i}
\label{eq:tensorterms}
\end{align}
terms in the tensorial basis. The summation variable $i$ counts the number of $g^{\mu \nu}$ in the corresponding terms. Some values of $N_{\text{terms}}$ are listed in table~\ref{tab:Nterms}.

\begin{table}
\centering
\begin{tabular}{|c|cccccc} \hline
$E \backslash \tau$ & 1 & 2 & 3 & 4 & 5 & 6 \\ \hline

0 &  0 & 1 & 0 & 3 & 0 & 15 \\

1 &  1 & 2 & 4 & 10 & 26 & 76 \\

2 &  2 & 5 & 14 & 43 & 142 & 499 \\

3 &  3 & 10 & 36 & 138 & 558 & 2364 \\

4 &  4 & 17 & 76 & 355 & 1724 & 8671 \\

5 &  5 & 26 & 140 & 778 & 4450 & 26140 \\

6 &  6 & 37 & 234 & 1515 & 10026 & 67731 \\
\end{tabular}
\caption{The number of terms in the tensorial basis for various values of $E$ and of the tensorial degree $\tau$. The numbers follow from eq.~\eqref{eq:tensorterms}.}
\label{tab:Nterms}
\end{table}

There can be more than one remaining dot-product for a loop, and an example of this would be $N(k) = (k \cdot p_3) (k \cdot p_4)$ in the example above. In that case $p_3$ and $p_4$ would have distinct $\eta$s, but that fact does not change much in the above discussion, the tensorial expansion of $p_3^{\mu} p_4^{\nu}$ would for instance have the same number of terms, non-spurious as well as spurious, as the expansion of $p_3^{\mu} p_3^{\nu}$ given by eq.~\eqref{eq:p3sq}.

For the loop-by-loop Baikov parametrization, the above algorithm should be applied at each loop at a time in the same order as the parametrization order, giving replacement rules for all combinations of remaining dot products. That is the idea behind the implementation in \texttt{BaikovPackage} as \texttt{MakeExtraDPRules} discussed in sec.~\ref{sec:features}. A different approach to the same problem is discussed in app.~\ref{app:recursive}, and that approach was used as a check for the expression generated with the method discussed here.

From the values listed in table~\ref{tab:Nterms} we see that the size of the system that has to be solved grows significantly with $E$ and even more so with $\tau$. For that reason the algorithm presented here should not be considered the last word on this question, it is merely intended as a proof of concept, showing (hopefully once and for all) that the presence of remaining dot products in the numerator is in no way an insurmountable obstacle for the loop-by-loop Baikov representation.

\section{The recursive structure}
\label{app:recursive}

In app.~\ref{app:derivation} we showed how to obtain the standard Baikov representation starting from loop-by-loop. It is, however, also possible to go the other way, from standard to loop-by-loop, by integrating the extra variables out one by one. This is explained in detail in refs.~\cite{Chen:2022lzr, Jiang:2023qnl} and this appendix will summarize the discussion there.

Integration variables that belong to the standard Baikov representation, but not to loop-by-loop, will always contain a scalar product of the form $k {\cdot} q$ (where $k$ is a loop momentum and $q$ is different from $k$) which is not present in other Baikov variables. Introducing the $n \times n$ Gram matrix $M = G(k, \ldots, q)$ of which $\mathcal{B}$ is the determinant, we may sort its arguments such that that scalar product appears as
\begin{align}
M_{1, n} = M_{n, 1} = \lambda x + \kappa
\end{align}
where $x$ is the Baikov variable that we want to integrate out, and where the other entries of $M$ are independent of $x$.

Let us now define the \textit{minor} notation, in which $A^{\alpha}_{\beta}$ is the determinant of the matrix formed out of the rows of the matrix $A$ indexed by $\alpha$ and the columns indexed by $\beta$.

$\mathcal{B}$ will be quadratic in $x$ so we may write it as
\begin{align}
\mathcal{B} = M_{1,\ldots, n}^{1,\ldots, n} = a x^2 + b x + c
\end{align}
and we can now identify
\begin{align}
a = - \lambda^2 M_{2,\ldots, n{-}1}^{2,\ldots, n{-}1} \,,\qquad b = \pm 2 \lambda M_{2,\ldots, n}^{1, \ldots, n{-}1} \big|_{x=0} \,,\qquad c = M_{1, \ldots, n}^{1, \ldots, n} \big|_{x=0} \,.
\end{align}
In particular we get for the discriminant
\begin{align}
b^2 - 4 a c \; = \; 4 \lambda^2 \left( \big( M_{2,\ldots, n}^{1, \ldots, n{-}1} \big)^2 + M_{2,\ldots, n{-}1}^{2,\ldots, n{-}1} M_{1, \ldots, n}^{1, \ldots, n} \right) \Big|_{x=0}
\end{align}
We will now use a form of \textit{Sylvester's determinant identity} known as the \textit{Desnanot--Jacobi identity} which for a generic $n \times n$ matrix $A$ states
\begin{align}
A_{1, \ldots, n{-}1}^{1, \ldots, n{-}1} A_{2, \ldots, n}^{2, \ldots, n} \; = \; A_{1, \ldots, n{-}1}^{2, \ldots, n} A_{2, \ldots, n}^{1, \ldots, n{-}1} + A_{2, \ldots, n{-}1}^{2, \ldots, n{-}1} A_{1, \ldots, n}^{1, \ldots, n}
\end{align}
Using also the facts that $M$ is symmetric and that $M_{1, \ldots, n{-}1}^{1, \ldots, n{-}1}$ and $M_{2, \ldots, n}^{2, \ldots, n}$ are independent of $x$ allows us to rewrite the discriminant as
\begin{align}
b^2 - 4 a c \; = \; 4 \lambda^2 M_{1, \ldots, n{-}1}^{1, \ldots, n{-}1} M_{2, \ldots, n}^{2, \ldots, n}
\end{align}

We can now use the general result
\begin{align}
\int_{r_1}^{r_2} \! ( a x^2 + bx + c )^{\gamma} \id x \; = \; \frac{\sqrt{\pi}}{2^{1 + 2 \gamma}} \frac{\Gamma(1 + \gamma)}{\Gamma(\tfrac{3}{2} + \gamma)} \, (-a)^{-1 - \gamma} \, (b^2 - 4 a c)^{\tfrac{1}{2} + \gamma}
\label{eq:recintbasic}
\end{align}
(where $r_1$ and $r_2$ are the two roots of the quadratic polynomial) to write the integration over the Baikov integrand as
\begin{align}
\int_{r_1}^{r_2} \! \mathcal{B}^{\gamma} \id x \; = \; \frac{\sqrt{\pi}}{\lambda} \, \frac{\Gamma(1 + \gamma)}{\Gamma(\tfrac{3}{2} + \gamma)} \, \mathcal{E}_1^{-1 - \gamma} \, \mathcal{B}_1^{\tfrac{1}{2} + \gamma} \, \mathcal{B}_2^{\tfrac{1}{2} + \gamma}
\label{eq:recursive}
\end{align}
where we have identified
\begin{align}
\mathcal{B}_1 = M_{1, \ldots, n{-}1}^{1, \ldots, n{-}1} \,,\qquad \mathcal{E}_1 = M_{2, \ldots, n{-}1}^{2, \ldots, n{-}1} \,,\qquad \mathcal{B}_2 = M_{2, \ldots, n}^{2, \ldots, n} \,.
\end{align}
Using this expression recursively for each extra scalar product, we may go from the standard to the loop-by-loop Baikov representation.\\

As an example, let us look at the doublebox with
\begin{align}
P_1 &= k_1^2 \,, & P_2 &= (k_1-p_1)^2 \,, & P_3 &= (k_1-p_1-p_2)^2 \,, & P_4 &= (k_2-p_1-p_2)^2 \,, & P_5 &= (k_2 + p_4)^2 \,, \nonumber \\
P_6 &= k_2^2 \,, & P_7 &= (k_1-k_2)^2 \,, & P_8 &= (k_2-p_1)^2 \,, & P_9 &= (k_1 + p_4)^2 \,.
\end{align}
Here we have $P_1$--$P_7$ being the genuine propagators of the doublebox. For the loop-by-loop parametrization (starting with the $k_1$-loop) will need $P_8$ as an ISP, while standard Baikov will need both $P_8$ and $P_9$. It is thus $x_9=P_9$ that we will have to integrate out when going from standard Baikov to loop-by-loop. We may write the scalar product only present in $P_9$ as $k_1 \cdot p_4 = \lambda x_9 + \kappa$ where $\lambda = \tfrac{1}{2}$ and $\kappa = -\tfrac{1}{2} x_1$. For standard Baikov we get from eq.~\eqref{eq:standard}
\begin{align}
I_{\text{doublebox}} &= \frac{2^{-7} \, \pi^{-7/2} \, \mathcal{E}^{(4-d)/2}}{\Gamma((d-4)/2) \Gamma((d-3)/2)} \int_{\mathcal{C}} \frac{ x_8^{-a_8} \; \mathcal{B}^{(d{-}6)/2} }{x_1^{a_1} \cdots x_{7}^{a_{7}}} \id^9 x
\end{align}
Applying the integration of eq.~\eqref{eq:recursive} to $x_9$, and identifying $\mathcal{E}$ with $\mathcal{E}_2$, we get
\begin{align}
I_{\text{doublebox}} &= \frac{2^{-6} \, \pi^{-3} \, \mathcal{E}_2^{(4-d)/2}}{\Gamma^2((d-3)/2)} \int_{\mathcal{C}} \frac{ x_8^{-a_8} \; \mathcal{B}_2^{(d{-}5)/2} \mathcal{E}_1^{(4{-}d)/2} \mathcal{B}_1^{(d{-}5)/2} }{x_1^{a_1} \cdots x_{7}^{a_{7}}} \id^8 x
\end{align}
which is exactly the loop-by-loop expression that would follow from eq.~\eqref{eq:loop-by-loop}.\\

We might consider what would happen if the variable that was integrated out was present in the numerator of the original integral, as also discussed in app.~\ref{app:remaining}. In that case the integrand of eq.~\eqref{eq:recintbasic} would contain an extra polynomial in $x$. The integral of the various monomials therein can be written as
\begin{align}
& \int_{r_1}^{r_2} \!\!\! x^n \, ( a x^2 + bx + c )^{\gamma} \id x \; = \; \frac{\sqrt{\pi}}{2^{1 + 2 \gamma}} \frac{\Gamma(1 + \gamma)}{\Gamma(\tfrac{3}{2} + \gamma)} \, (-a)^{-1 - \gamma} \, (b^2 - 4 a c)^{\tfrac{1}{2} + \gamma} \, X_n
\label{eq:recintn}
\end{align}
where $X_n$ can be expressed in closed form as
\begin{align}
X_n &= \bigg( \frac{-b}{2a} \bigg)^{\!n} \, {}_2 F_1 \! \left( \tfrac{-n}{2} , \tfrac{1-n}{2} , \tfrac{3}{2} {+} \gamma ; \tfrac{b^2-4ac}{b^2} \right)
\end{align}
That expression can, however, look misleading since that hypergeometric function is rational whenever $n$ is a non-negative integer. For that reason we will also list some specific values, which are
\begin{align}
X_0 = 1 \,,\qquad X_1 = \frac{-b}{2a} \,,\qquad X_2 = \frac{b^2 (\gamma{+}2) - 2 a c}{2 a^2 (2 \gamma{+}3)} \,,\qquad X_3 = \frac{6 a b c - b^3 (\gamma{+}3)}{4 a^3 (2 \gamma{+}3)} \,, \nonumber
\end{align}
\vspace{-4mm}
\begin{align}
X_4 = \frac{12 a^2 c^2-12 a b^2 c (\gamma{+}3)+b^4 (\gamma{+}3) (\gamma{+}4)}{4 a^4 (2 \gamma{+}3) (2 \gamma{+}5)} \,,\quad 
X_5 = \frac{-60 a^2 b c^2+20 a b^3 c (\gamma{+}4) - b^5 (\gamma{+}4) (\gamma{+}5)}{8 a^5 (2 \gamma{+}3) (2 \gamma{+}5)} \,, \nonumber
\end{align}
\vspace{-4mm}
\begin{align}
X_6 = \frac{-120 a^3 c^3+180 a^2 b^2 c^2 (\gamma{+}4)-30 a b^4 c (\gamma{+}4) (\gamma{+}5)+b^6 (\gamma{+}4) (\gamma{+}5) (\gamma{+}6)}{8 a^6 (2 \gamma{+}3) (2 \gamma{+}5) (2 \gamma{+}7)} \,.
\end{align}
Eq.~\eqref{eq:recintn} provides an alternative approach to that of app.~\ref{app:remaining} for deriving loop-by-loop Baikov parametrizations when the extra scalar products are present in the numerator, and in fact we used this equation to double-check the expressions given there.

The recursive approach discussed in this appendix is the starting point for the investigation into connections between the Baikov representation and canonical differential equations and \textit{symbol letters}, which is performed in refs.~\cite{Chen:2022lzr, Jiang:2023qnl}.

\newpage

\bibliographystyle{JHEP}
\bibliography{biblio}

\end{document}